\documentclass[aps,floats,showpacs,nofootinbib, twocolumn]{revtex4-1} 

\usepackage{amssymb}
\usepackage{amsmath}

\usepackage{sans}

\usepackage{graphicx}

\usepackage{graphics}
\usepackage{dcolumn}
\usepackage{color}
\usepackage{rotate}
\usepackage{fancyhdr}
\usepackage{hyperref}
\usepackage{indentfirst}

\usepackage{umoline}
\usepackage{ulem}

\newcommand{\apjl}{Astrophys.\ J.\ Lett.\  }

\newcommand{\mnras}{Mon.\ Not.\ R.\ Astron.\ Soc.\ }

\usepackage{physics}
\usepackage{booktabs}
\usepackage{fontawesome}

\def\beq{\begin{equation}}
\def\eeq{\end{equation}}
\def\be{\begin{eqnarray}}
\def\ee{\end{eqnarray}}
\def\ba{\begin{eqnarray}}
\def\ea{\end{eqnarray}}
\def\no{\nonumber}

\newcommand{\iu}{\mathrm{i}}
\newcommand{\e}{\mathrm{e}}

\definecolor{darkred}{rgb}{.743,0,0}

\def\n02b{$0\nu\beta\beta$}
\def\n02bphi{$0\nu\beta\neta\phi$}
\def\lsim{\mathrel{\rlap{\lower4pt\hbox{\hskip1pt$\sim$}}
    \raise1pt\hbox{$<$}}}         
\def\gsim{\mathrel{\rlap{\lower4pt\hbox{\hskip1pt$\sim$}}
    \raise1pt\hbox{$>$}}}         

\begin{document}
\title{Comments on the mass sheet degeneracy in cosmography analyses
}

\author{Luca Teodori}\email{luca.teodori@weizmann.ac.il}
\affiliation{Weizmann Institute, Department of Particle Physics and Astrophysics, Rehovot, Israel 7610001}

\author{Kfir Blum}\email{kfir.blum@weizmann.ac.il}
\affiliation{Weizmann Institute, Department of Particle Physics and Astrophysics, Rehovot, Israel 7610001}

\author{Emanuele Castorina}\email{emanuele.castorina@unimi.it}
\affiliation{Dipartimento di Fisica ‘Aldo Pontremoli’, Universita’ degli Studi di Milano, Via Celoria 16, 20133 Milan, Italy}

\author{Marko Simonovi\'c}\email{marko.simonovic@cern.ch}
\affiliation{Theoretical Physics Department, CERN,
1 Esplanade des Particules, Geneva 23, CH-1211, Switzerland}

\author{Yotam Soreq}\email{soreqy@physics.technion.ac.il}
\affiliation{Physics Department, Technion -- Israel Institute of Technology, Haifa 3200003, Israel}


\begin{abstract}
We make a number of comments regarding modeling degeneracies in strong lensing measurements of the Hubble parameter $H_0$. The first point concerns the impact of weak lensing associated with different segments of the line of sight. We show that external convergence terms associated with the lens-source and observer-lens segments need to be included in cosmographic modeling, in addition to the usual observer-source term, to avoid systematic bias in the inferred value of $H_0$. Specifically, we show how an incomplete account of some line of sight terms biases stellar kinematics as well as ray tracing simulation methods to alleviate the mass sheet degeneracy.
The second point concerns the use of imaging data for multiple strongly-lensed sources in a given system. 
We show that the mass sheet degeneracy is not fully resolved by the availability of multiple sources: some degeneracy remains because of differential external convergence between the different sources. 
Similarly, differential external convergence also complicates the use of multiple sources in addressing the approximate mass sheet degeneracy associated with a local (``internal") core component in lens galaxies. This internal-external degeneracy is amplified by the non-monotonicity of the angular diameter distance as a function of redshift. 
For a rough assessment of the weak lensing effects, we provide estimates of external convergence using the nonlinear matter power spectrum, paying attention to non-equal time correlators.
\end{abstract}

\maketitle
\tableofcontents
\section{Introduction}
Strong gravitational lensing of galaxies probes the mass distribution in lens objects and the background cosmology~\cite{Blandford:1991xc,Schneider:1992,Kochanek:2004ua,Treu:2010uj}. Imaging data combined with gravitational time delays of quasars and supernovae could allow a determination of the Hubble parameter $H_0$~\cite{10.1093/mnras/128.4.307,Suyu:2012aa,Treu:2016ljm,Suyu:2016qxx,Grillo:2020yvj}. Subject to simplifying assumptions on the mass profile of lens galaxies, a handful of systems with quasar time delays were enough for measurements of $H_0$~\cite{Rusu:2019xrq,Birrer:2018vtm,DES:2019fny,Chen:2019ejq, Wong:2019kwg,Millon:2019slk} that were widely used for tests of the $\Lambda$CDM model~\cite{SHOES,Verde:2019ivm,DiValentino:2021izs}. 
The prospects for time delay cosmography will make a leap with the advent of various surveys~\cite{DES:2018whv,SKA:2018ckk,Euclid:2019clj} and notably the  LSST~\cite{LSSTScience:2009jmu}, that will discover thousands of lensed quasars and dozens of lensed supernovae, bringing the number of strongly lensed quasars with time delay measurements to hundreds~\cite{Oguri:2010ns,Liao:2014cka,Dobke:2009bz}; and with the JWST~\cite{Gardner:2006ky}, that will sharpen constraints on lens stellar kinematics~\cite{Birrer:2020jyr,Yildirim:2021wdd}. 

While observations become numerous and precise, systematic degeneracies are a well known limiting factor in the interpretation of lensing data~\cite{1991ApJ...373..354K,Kochanek:2002rk,Liesenborgs:2012pu,Schneider_2013,Kochanek:2020crs}.  
For example, relaxing some of the simplifying assumptions made in~\cite{Rusu:2019xrq,Birrer:2018vtm,DES:2019fny,Chen:2019ejq, Wong:2019kwg,Millon:2019slk}, a possible tension between the value of $H_0$ inferred via lensing and via large-scale structure (LSS) analyses~\cite{Akrami:2018vks,DAmico:2019fhj,Ivanov:2019pdj} may be replaced by a core feature in the lenses~\cite{Blum:2020mgu,Birrer:2020tax}. It is clear that a careful account of  modeling degeneracies will be crucial to take advantage of the progress in observations. 

In this paper we comment on certain modeling degeneracies that affect the connection of imaging, time delay, and kinematics with physical information on lens profiles and cosmology. Several aspects of our discussion have been considered in the past at various levels of detail (see, e.g.~\cite{Falco1985,1991ApJ...373..354K,Kochanek:2002rk,Schneider_2013}, and notably the discussion in~\cite{Birrer:2020tax}). However, as we show, recent cosmography campaigns still do not account for the degeneracy in full.

The outline of the paper is as follows. In Sec.~\ref{s:msdweak} we review cosmological weak lensing effects~\cite{Bartelmann:1999yn,Kaiser:1992ps}, that are intertwined with the strong lensing reconstruction problem via the mass sheet degeneracy (MSD). 

In Sec.~\ref{s:kinmsd} we show a limitation in using kinematics data to alleviate the MSD. The problem is that weak lensing entails three distinct effects, coming from the source-observer segment, the source-lens segment, and the lens-observer segment of the line of sight (LOS). Omitting shear for a moment, these effects are summarised by  three convergence terms: $\kappa^{\rm s}$, $\kappa^{\rm ls}$, and $\kappa^{\rm l}$, respectively. In general, different combinations of $\kappa^{\rm s}$, $\kappa^{\rm ls}$, and $\kappa^{\rm l}$ enter into the bias in $H_0$ and into the interpretation of kinematics data. 
To ameliorate this ambiguity, imaging+kinematics analyses such as~\cite{Rusu:2019xrq,Birrer:2018vtm,DES:2019fny,Chen:2019ejq, Wong:2019kwg,Millon:2019slk,Birrer:2020tax,Yildirim:2021wdd} should introduce nuisance parameters for $\kappa^{\rm l}$, in addition to $\kappa^{\rm s}$.

In Sec.~\ref{s:raymsd} we discuss the use of ray tracing simulations to obtain an observationally-informed theoretical prior on the weak lensing correction. 
We note that accounting for the full bias in $H_0$ requires that the ray tracing be used to extract all of $\kappa^{\rm s}$, $\kappa^{\rm ls}$, and $\kappa^{\rm l}$. Existing analyses~\cite{Rusu:2019xrq,Birrer:2018vtm,DES:2019fny,Chen:2019ejq, Wong:2019kwg,Millon:2019slk} neglected the $\kappa^{\rm ls}$ and $\kappa^{\rm l}$ terms, possibly resulting in residual bias to their inferred value of $H_0$.

In Sec.~\ref{sec:modeling_without_kappaext} we consider systems with multiple sources. This is a timely problem because analyses of multiple sources in cluster lens systems are advancing~\cite{Grillo:2020yvj}, and the situation also occurs in some galaxy lenses~\cite{Collett:2014ola,DES:2019fny}, where we can expect significant observational progress with the advent of new surveys~\cite{Oguri:2010ns, Gavazzi:2008aq,Liao:2014cka}. 
We show that the MSD associated with weak lensing is not resolved by multiple sources, and clarify what imaging data does measure: a certain difference-of-differences of convergence terms. This combination of convergence terms is not the same one that enters the $H_0$ reconstruction problem. To mitigate the MSD, theoretical estimates of the weak lensing effect must be input to the analysis, similarly (although not precisely the same) to the way it needs to be input in systems with a single source. Multi-source analyses such as  Ref.~\cite{Grillo:2020yvj} should be adjusted to include this effect.

In Sec.~\ref{ss:msdcore} we consider the so-called internal MSD. Specifically, we are thinking of the impact of a sub-dominant core component in the density profile of lens galaxies, which could act as an approximate version of the MSD~\cite{Schneider_2013,Blum:2020mgu}. We show that imaging data by itself cannot distinguish a core deformation of the intrinsic lens model, from an adjustment of weak external convergence. Interestingly, this statement remains true even when multiple sources are available. Introducing theoretical estimates of weak lensing, it should indeed become possible to identify a core effect if the magnitude of the convergence term associated with the core is parametrically larger than that expected theoretically from weak lensing. However, we point out an important dilution factor that makes this distinction potentially difficult.
In Sec.~\ref{sec:real_systems} we estimate the effect for two sample systems. 
%

A related discussion of the multi-source MSD was given in~\cite{McCully:2013fga,Schneider:2014vka,Schneider:2014ifa}. The main difference between that work and ours here, is that~\cite{McCully:2013fga,Schneider:2014vka,Schneider:2014ifa} considered the role of intermediate sources as additional deflectors, that must be modeled separately from the main lens, and that exhibit a residual multi-lens version of the MSD. We comment on this point in App.~\ref{sec:deriveLSS}. It does not replace our discussion, but adds another layer of complexity in the modeling.


We summarise in Sec.~\ref{s:sum}. 
In App.~\ref{sec:deriveLSS} we give a brief derivation of the weak lensing correction in strong lensing systems. 
In App.~\ref{sec:kappa_variance} we provide estimates of weak external convergence using the nonlinear matter power spectrum and paying attention to non-equal time correlators that arise due to projection. We use these computations for a rough assessment of the  effect. This is enough for illustrating our main points in this paper, although direct weak lensing surveys or ray tracing techniques, specifically designed to match the bias of the field containing individual strong lensing systems~\cite{Keeton:1996tq,Holder:2002hq,Dalal:2004as,Momcheva:2005ex,Suyu:2009by,2017MNRAS.467.4220R,2018ApJ...867..107W,H0LiCOWX2019,2020MNRAS.498.1406T}, are probably mandatory for more accurate analyses.

\section{Recap: weak lensing and mass sheet degeneracy in strong lensing analyses}\label{s:msdweak}
Consider a gravitational lens system with $ N$ sources, located at redshifts $ z_i $, $ i = 1,\ldots, N $. The deflection angle caused by the lens (main deflector) relative to source $ i $ reads
\be\label{eq:ams}\vec\alpha_i(\vec\theta)&=&\frac{1}{\pi}\int d^2\theta'\frac{\vec\theta-\vec\theta'}{|\vec\theta-\vec\theta'|^2}\kappa_i(\vec\theta') .
\ee
Here $\kappa_{i}(\vec\theta)$ is the convergence,  
\be\kappa_i(\vec\theta)&=& \frac{\Sigma\left(d_{\rm A}(0,z_{\rm l})\vec\theta\right)}{\Sigma_{\rm crit}(z_{\rm l},z_i)}, \ee
 $\Sigma(\vec x)$ is the surface mass density of the lens computed at proper position $\vec x$ transverse to the observer-lens line of sight, $\Sigma_{\rm crit}(z_{\rm l},z_i)$ is the critical density (we use natural units with $c=1$),
\be
\label{eq:Sc}\Sigma_{\rm crit}(z_{\rm l},z_i)&=&\frac{1}{4\pi G}\frac{d_{\rm A}(0,z_i)}{d_{\rm A}(0,z_{\rm l})\,d_{\rm A}(z_{\rm l},z_i)},
\ee
 $d_{\rm A}(z_{\rm o},z_{\rm e})$ is the angular diameter distance from an emitter at redshift $z_{\rm e}$ to an observer at $z_{\rm o}$, and $z_i,\,z_{\rm l}$ are the redshifts of the $ i $-th source and of the lens, respectively.
 Notice that we can write 
 \be\label{eq:aia1}
 \vec\alpha_i(\vec\theta) = C_i \vec\alpha_1(\vec\theta) \ , \ C_i := \frac{d_{\rm A}(0,z_{1})\,d_{\rm A}(z_{\rm l},z_{i},)}{d_{\rm A}(0,z_{i})\,d_{\rm A}(z_{\rm l}, z_{1})}.
 \ee
That is, the deflection angle affecting the $i$-th source is a scaled version of the deflection angle affecting the $1$st source. When we discuss the internal lens model in what follows it would be convenient to highlight $\vec\alpha_1$, understanding that $\vec\alpha_i$ follows by Eq.~\eqref{eq:aia1}. 

In writing $\vec\alpha(\vec\theta)$ we think of the main deflector as a localised concentration of mass (localised compared with cosmological distances), assuming that $\alpha(\vec\theta)\to0$ for $|\vec\theta|$ much larger than the Einstein angle of the system, $|\vec\theta_{\rm E}|$, defined via\footnote{The definition of $\vec\theta_{\rm E}$ in Eq.~(\ref{eq:aEtE}) applies for axisymmetric lenses, but may not apply for arbitrary lens mass distributions. This subtlety is not important for our analysis.}
\be\label{eq:aEtE}
\vec\alpha(\vec\theta_{\rm E})&=&\vec\theta_{\rm E}.
\ee

Weak lensing from large scale structure in the intervening space between the sources, the lens, and the observer, modifies the lens equation by introducing external convergence and shear. These modifications must be taken into account in lensing analyses~\cite{McCully:2016yfe}. In the tidal approximation, the lens equation becomes~\cite{1991ApJ...380....1M,Kaiser:1992ps,Bar-Kana:1995qyu,Keeton:1996tq,McCully:2013fga,Schneider:2014vka,Fleury:2021tke} (see also App.~\ref{sec:deriveLSS})
\begin{align} \label{weakLens_original}
\begin{aligned}
\vec{\beta}_i &= (1- \kappa_i^{\rm s}) (\mathbb{I} + \Gamma_i^{\rm s}) \vec{\theta}\\ 
-&(1- \kappa_i^{\rm ls}) (\mathbb{I} + \Gamma_i^{\rm ls}) C_i \vec{\alpha}_1( (1- \kappa^{\rm l}) (\mathbb{I} + \Gamma^{\rm l}) \vec{\theta}) ,
\end{aligned}
\end{align}
where $ \kappa^{\rm r}_i $ are external convergence factors for source $ i $, 
\be
\Gamma_i^{\rm r} = 
-\begin{pmatrix}
 \gamma^{\mathrm{r}, i}_1 & \gamma^{\mathrm{r}, i}_2 \\
\gamma^{\mathrm{r}, i}_2 &  -\gamma^{\mathrm{r}, i}_1
\end{pmatrix} 
\ee
is the reduced shear matrix, and the superscript $\rm{r} = \rm{l}, \rm{s}, \rm{ls}$ indicates observer-lens, observer-source, and lens-source lines of sight. 

\begin{figure}
    \centering
    \includegraphics[scale=0.4]{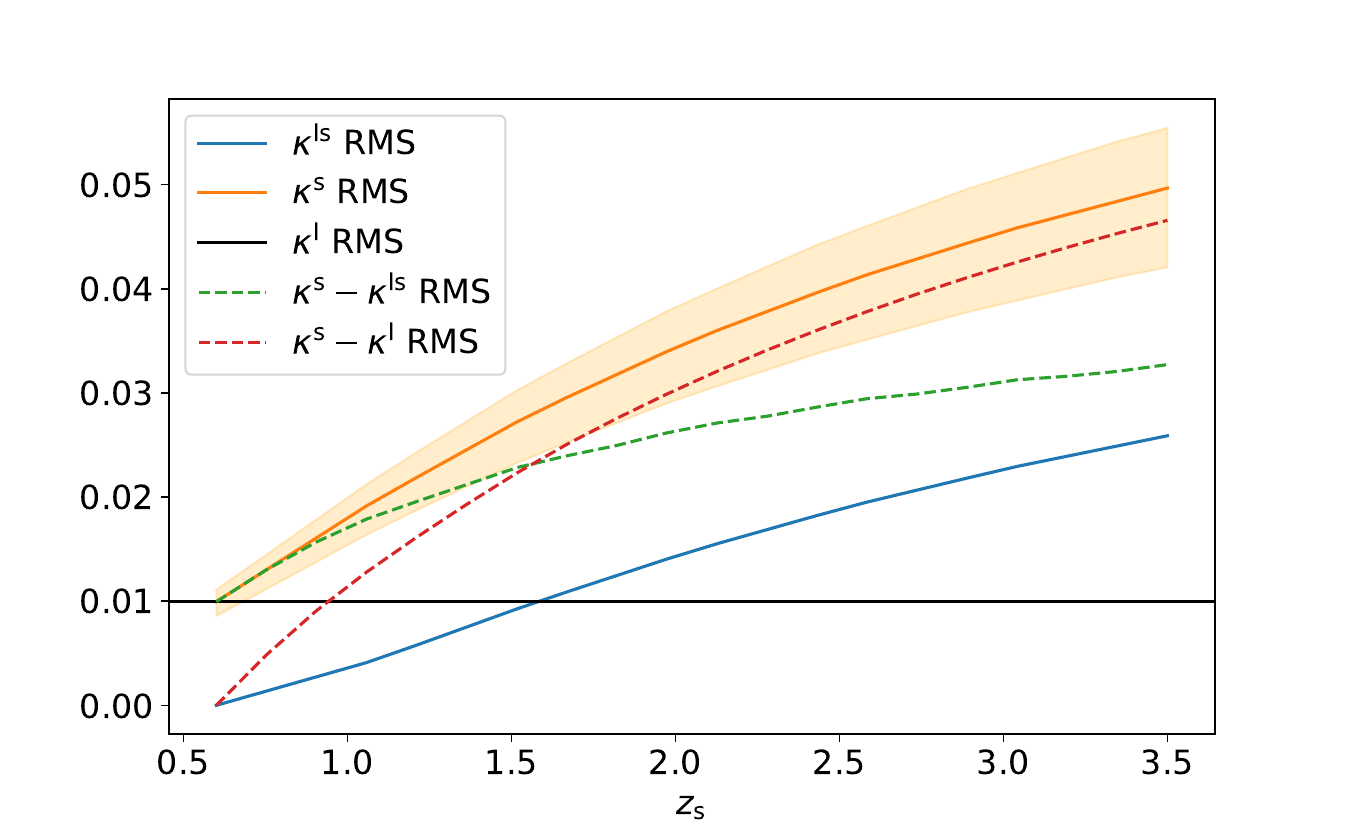}
    \caption{RMS external convergence terms, for lens redshift $z_{\rm l}=0.59$, presented as functions of the source redshift $z_{\rm s}$. The orange band around $\kappa^{\rm s}$ shows a rough estimate of the theoretical uncertainty, obtained by varying the cutoff of the matter power spectrum calculation from $k_{\rm cutoff}=5~{\rm Mpc}^{-1}$ to $20~{\rm Mpc}^{-1}$; the default in the calculation is $10~{\rm Mpc}^{-1}$. Modifying $k_{\rm cutoff}$ has a similar effect on the other weak convergence terms in the plot. Details of the calculation are given in App.~\ref{sec:kappa_variance}. Note that $\kappa^{\rm s}, \kappa^{\rm ls}, \kappa^{\rm l}$ are statistically independent (albeit correlated) cosmological random variables; thus, for example, the RMS value of $\kappa^{\rm s}-\kappa^{\rm l}$ is not simply shifted by a constant from the RMS value of $\kappa^{\rm s}$, even though the RMS of $\kappa^{\rm l}$ is a constant (given that the plot is done at constant fixed $z_{\rm l}$). Note that this plot is not expected to be accurate beyond the $\mathcal{O}(1)$ level. More accurate results would require ray tracing techniques to capture bias from excess of structure along the  LOS~\cite{Keeton:1996tq,Holder:2002hq,Dalal:2004as,Momcheva:2005ex,Suyu:2009by,2017MNRAS.467.4220R,2018ApJ...867..107W,H0LiCOWX2019,2020MNRAS.498.1406T}. Code: \href{https://github.com/lucateo/Comments_MSD/blob/main/Notebooks/delta_kappa_nonlinear.ipynb}{\faGithub}.}
    \label{fig:kappa_ls_s_l}
\end{figure}

Compared with the internal convergence $\kappa_i$, which is of order unity near the Einstein angle $\kappa_i(\vec\theta_{{\rm E},i})=\mathcal{O}(1)$, the weak lensing terms are small, typically in the range $|\gamma^{\rm r,i}|,\,|\kappa^{\rm r}_i|\sim0.01-0.1$. In App.~\ref{sec:kappa_variance} we estimate their magnitude; a typical result is illustrated in Fig.~\ref{fig:kappa_ls_s_l}. We show the root mean square (RMS) values of $\kappa^{\rm l,s,ls}$, which are cosmological random variables. The shear terms $\gamma^{\rm l,s,ls}_{1,2}$ scale similarly. 
For coherence with the tidal approximation, in the following we will mostly keep first order in $\kappa^{{\rm r},i} $, $\gamma^{{\rm r}, i}_{1,2}$. We assume that the large-scale structure producing the weak lensing is distributed over cosmological scales $\gtrsim1$~Mpc (compared with the galactic scale $\ll1$~Mpc of the primary lens that produces $\vec\alpha_i$), thus the weak lensing terms are approximated as constants over the angular range containing the strong lensing image information.

For simplicity of notation, we define
\begin{equation}\label{eq:Mdef}
(1- \kappa_i^{\rm r}) (\mathbb{I} + \Gamma_i^{\rm r}) \simeq \mathbb{I} - \left(\mathbb{I}\kappa_i^{\rm r}- \Gamma_i^{\rm r}\right) =: \mathbb{I} - M^{\rm r}_i .
\end{equation}
Note that $M^{\rm s}_i$ and $M^{\rm ls}_i$ carry the source label $i$, while $M^{\rm l}$ is common to all sources.  
With this notation, we can write a weak lensing-modified lens equation as
\be\label{eq:lensweak}\vec\beta_i&=&\vec\theta-\vec{\tilde\alpha}_i(\vec\theta),\\
\label{tilde_alpha}
\vec{\tilde{\alpha}}_i(\vec{\theta}) &=& (\mathbb{I} - M^{\rm ls}_{i} )C_i\vec{\alpha}_1((\mathbb{I} - M^{\rm l})\vec\theta) +  M_i^{\rm s} \vec\theta.
\ee
The modified deflection angle $\vec{\tilde\alpha}$  contains a mixture of terms, some local to the lens and some coming from weak lensing. Thus $\vec{\tilde\alpha}(\vec\theta)$, in general, does not decay at large $|\theta|$; instead, it satisfies $\vec{\tilde\alpha}_i(\vec\theta)\to M_i^{\rm s}\vec\theta$.

The time delay between images $A$ and $B$ (associated, e.g., to a time-variable quasar) of source $i$ is \cite{Kovner:1987,Bar-Kana:1995qyu,Schneider:1997bq,Schneider:2014vka} (see also App.~\ref{sec:deriveLSS})
\be \label{eq:DtAB}
\Delta t^i_{AB} &=&D^i_{\rm dt}\, \Delta\tau^i_{AB},\\
\label{eq:DtauAB}\Delta\tau^i_{AB}&=&\frac{1}{2}\vec\theta_A^T\left(\mathbb{I}-M_i^{\rm s}-M^{\rm l}+M_i^{\rm ls}\right)\vec\theta_A\no\\
&-&\vec\beta^T\left(\mathbb{I}-M^{\rm l}+M_i^{\rm ls}\right)\vec\theta_A-\psi_i((\mathbb{I}-M^{\rm l})\vec\theta_A)\no\\
&-&\{A\leftrightarrow B\}.
\ee
Here $D^i_{\rm dt}$ is the time-delay distance~\cite{1975ApJ...195L..11C},
\be
D^i_{\rm dt} := (1+z_{\rm l} ) \frac{d_{\rm A}(0,z_{\rm l}) d_{\rm A}(0,z_i)}{d_{\rm A}(z_{\rm l},z_i)} \propto \frac{1}{H_0}  ,
\ee
and $\psi_i(\vec\theta)=C_i\psi_1(\vec\theta)$ 
is the intrinsic lensing potential, defined via $\vec\nabla\psi_i(\vec\theta) =\vec{\alpha}_i(\vec\theta)$. In this analyses we do not explore the possibility of obtaining time-delay data for more than one source. Thus, we will drop the source index $^i$ on $\Delta\tau_{AB}$.

The MSD affecting the lensing reconstruction problem~\cite{1985ApJ...289L...1F}  
is usually represented by replacing, in Eqs.~\eqref{eq:lensweak} and~\eqref{tilde_alpha},
\be \label{msd_transform1}
\vec{\beta}_i &\longmapsto& \vec\beta_{i}^\lambda=\lambda \vec{\beta}_i,\\
\label{msd_transform2}\vec{\tilde{\alpha}}_i(\vec\theta) &\longmapsto& \vec{\tilde\alpha}_{i}^\lambda(\vec\theta)=\lambda \vec{\tilde{\alpha}}_i(\vec\theta) + (1- \lambda)\vec{\theta} ,
\ee
where $\lambda$ is an arbitrary real parameter. (More general degeneracies exist~\cite{Schneider:2013wga,Unruh:2016adf,Wertz:2017dxa}, but for our main points it is enough that we restrict ourselves to Eqs.~(\ref{msd_transform1}-\ref{msd_transform2}).) 
Image coordinates $\vec\theta$ and magnification ratios are invariant under Eqs.~(\ref{msd_transform1}-\ref{msd_transform2}). 
However, time delays are affected, and therefore, so is the inference of $H_0$. 

Eqs.~(\ref{msd_transform1}-\ref{msd_transform2}) imply a degeneracy in the modeling of weak lensing data, coupled with a reparameterization of the model describing the ``intrinsic" deflection angle $\vec\alpha_1$. Infinitely many different reparameterizations of $M^{\rm r}_i$ and $\vec\alpha_1$ can produce Eqs.~(\ref{msd_transform1}-\ref{msd_transform2}).
In considering these possibilities we assume that lens and source redshifts are measured perfectly, so the cosmological functions $C_i$ are known without appreciable uncertainty (given a cosmological model).

Because of the inhomogeneous term $(1-\lambda)\vec\theta$ in Eq.~\eqref{msd_transform2}, it is natural to associate the MSD with a reinterpretation of the inhomogeneous observer-source weak lensing term in Eq.~\eqref{tilde_alpha}, via $M_i^{\rm s}\,\longmapsto\,M_i^{\rm s,\lambda}=\lambda_{\rm s} M_i^{\rm s}+(1-\lambda_{\rm s})\mathbb{I}$. However, the interpretation is coupled to additional degeneracies with $M^{\rm ls}_i$ and $M^{\rm l}$. 
It is convenient to parameterize the combined degeneracy by allowing $M^{\rm ls}_i$ and $M^{\rm l}$ to also be adjusted, alongside an adjustment of the intrinsic lens model and the modeled source coordinates:
\be\label{eq:Mmsds}
M_i^{\rm s}&\longmapsto&M_i^{\rm s,\lambda}=\lambda_{\rm s} M_i^{\rm s}+(1-\lambda_{\rm s})\mathbb{I},\\
\label{eq:Mmsdls}
M_i^{\rm ls}&\longmapsto&M_i^{\rm ls,\lambda}=\lambda_{\rm ls} M_i^{\rm ls}+(1-\lambda_{\rm ls})\mathbb{I},\\
\label{eq:Mmsdl}M^{\rm l}&\longmapsto&M^{\rm l,\lambda}=\lambda_{\rm l} M^{\rm l}+(1-\lambda_{\rm l})\mathbb{I},\\
\label{eq:bmsd}\vec{\beta}_i &\longmapsto& \vec\beta_{i}^\lambda=\lambda_{\rm s} \vec{\beta}_i,\\
\label{eq:amsd}
\vec\alpha_1(\vec\theta)&\longmapsto&\vec\alpha_1^\lambda(\vec\theta)=\lambda_{\rm s}\,\lambda_{\rm ls}^{-1}\,\vec\alpha_1(\lambda_{\rm l}^{-1}\,\vec\theta),\\
\label{eq:psimsd}\psi_1(\vec\theta)&\longmapsto&\psi_1^\lambda(\vec\theta)=\lambda_{\rm s}\,\lambda_{\rm ls}^{-1}\,\lambda_{\rm l}\psi_1(\lambda_{\rm l}^{-1}\,\vec\theta).
\ee
Here $\lambda_{\rm s},\lambda_{\rm ls}$, and $\lambda_{\rm l}$ are independent parameters. Note that Eq.~(\ref{eq:Mmsds}) (for example) amounts to $\kappa^{\rm s}_i\,\longmapsto\,\kappa^{\rm s, \lambda}_i = \lambda_{\rm s} \kappa^{\rm s}_i + (1 - \lambda_{\rm s}),\;\;\Gamma^{\rm s}_i,\longmapsto\,\Gamma^{\rm s, \lambda}_i = \lambda_{\rm s} \Gamma^{\rm s}_i$. 

Inserting Eqs.~(\ref{eq:bmsd}-\ref{eq:psimsd}) into Eq.~(\ref{eq:DtauAB}), we see that the dimensionless time delay $\Delta\tau_{AB}$ of the transformed model changes according to (see~\cite{Birrer:2020tax} for an earlier discussion):
\be\Delta\tau_{AB}&\longmapsto&\Delta\tau_{AB}^\lambda=\lambda_{\rm s}\lambda^{-1}_{\rm ls}\lambda_{\rm l}\Delta\tau_{AB}.\ee
Thus, a readjustment of the lensing model according to Eqs.~(\ref{eq:Mmsds}-\ref{eq:bmsd}) entails a reinterpretation of the inferred value of $H_0$. Since $H_0$ is inferred from the measured time delays $\Delta t_{AB}$ and the model dimensionless time delay $\Delta\tau_{AB}$ via $H_0\propto\Delta\tau_{AB}/\Delta t_{AB}$, we have:
\be H_0&\longmapsto&H_0^\lambda=\lambda_{\rm s}\lambda^{-1}_{\rm ls}\lambda_{\rm l}H_0.\ee

We would like to emphasize that the availability of multiple sources does not, by itself, ameliorate the MSD: as far as imaging information is considered, the modeling degeneracy expressed by Eqs.~(\ref{eq:Mmsds}-\ref{eq:psimsd}) remains exact. It simply amounts to a simultaneous reinterpretation of the weak lensing variables affecting all of the sources. (The same conclusion, with a different version of the MSD and a discussion of intermediate sources as additional strong lenses for background sources, was reached in Refs.~\cite{McCully:2013fga,Schneider:2014vka}.) We return to this point in Sec.~\ref{sec:modeling_without_kappaext}.

In the absence of a direct measurement of weak lensing applicable to the field of view of the strong lensing system, the only way to ameliorate the MSD is by appealing to theoretical estimates of the magnitude of weak lensing variables. For example, a theoretical estimate of the expected possible magnitude of $\kappa_i^{\rm s}$, as shown in Fig.~\ref{fig:kappa_ls_s_l}, could constrain the conceivable range of $1-\lambda_{\rm s}$ in Eq.~(\ref{eq:Mmsds}): for some systems, an additive shift of order $|1-\lambda_{\rm s}|\approx0.1$ in $\kappa^{\rm s}_i$ may be difficult to justify from a cosmological point of view. In App.~\ref{sec:kappa_variance} we estimate some of these theoretical constraints.

\section{On the use of stellar kinematics to resolve the MSD}\label{s:kinmsd}
In an imaging analysis, if only a single source is available (say $i=1$), one can use Eqs.~(\ref{eq:Mmsds}-\ref{eq:psimsd}) with the choice 
\be
\lambda_{\rm s}&=&\frac{1}{1-\kappa^{\rm s}},\;\;\;\lambda_{\rm ls}\,=\,\frac{1}{1-\kappa^{\rm ls}},\;\;\;\lambda_{\rm l}\,=\,\frac{1}{1-\kappa^{\rm l}},\ee
to eliminate all of $\kappa^{\rm s},\,\kappa^{\rm ls},$ and $\kappa^{\rm l}$ from the modeling. For this reason, the task of extracting lensing information in imaging data is often performed ignoring external convergence
~\cite{Rusu:2019xrq,DES:2019fny,Chen:2019ejq, Wong:2019kwg,Millon:2019slk,Birrer:2018vtm}. (The details of how shear is modeled~\cite{Birrer:2016xku} will not be important for the discussion in this section.)

Suppose we denote the fit result for the ``intrinsic deflection angle" in such an analysis by $\vec\alpha^{\rm model}(\vec\theta)$. By ``eliminating external convergence from the equations", we mean that the fit looks for a deflection angle model $\vec\alpha^{\rm model}(\vec\theta)$ which goes to zero at large $|\vec\theta|$, possibly up to a uniform shear term $\Gamma^{\rm s}\vec\theta$. 
Eq.~\eqref{eq:amsd} implies that $\vec\alpha^{\rm model}(\vec\theta)$ is related to the true underlying physical intrinsic deflection angle by 
\be\label{eq:alphamodel} \vec\alpha^{\rm model}(\vec\theta)&=&\frac{1-\kappa^{\rm ls}}{1-\kappa^{\rm s}}\vec\alpha((1-\kappa^{\rm l})\vec\theta),\ee
where $\kappa^{\rm s,ls,l}$ are the true physical values of the weak lensing terms. Given a measurement of the physical image time delays, and deriving the dimensionless time delay $\Delta\tau_{AB}^{\rm model}$ from $\vec\alpha^{\rm model}$, one can extract an inferred result $H_0^{\rm model}$, which is related to the truth value $H_0$ by~\cite{Birrer:2020tax}
\be\label{eq:H0modcorrection} H_0^{\rm model}&=&\frac{1-\kappa^{\rm ls}}{(1-\kappa^{\rm s})(1-\kappa^{\rm l})}H_0.\ee
The usual challenge of the weak lensing MSD for cosmography is to constrain the correction factor $(1-\kappa^{\rm ls})/[(1-\kappa^{\rm s})(1-\kappa^{\rm l})]\approx1+\kappa^{\rm s}+\kappa^{\rm l}-\kappa^{\rm ls}$.

Stellar kinematics is sensitive to the intrinsic mass-per-radius ($M(R)/R$) of the lens, and can be used to partially resolve the MSD. 
Refs.~\cite{Rusu:2019xrq,Birrer:2018vtm,DES:2019fny,Chen:2019ejq, Wong:2019kwg,Millon:2019slk} used kinematics to constrain the MSD, but in these works, weak lensing was only parameterised in terms of $\kappa^{\rm s}$, omitting $\kappa^{\rm ls}$ and $\kappa^{\rm l}$. The omission of $\kappa^{\rm ls}$ and $\kappa^{\rm l}$ biases the inferred value of $H_0$. To explain this we consider a simplified scenario, where we can inspect the information content of imaging, time delays, and kinematics separately. 

Suppose that the intrinsic deflection angle of the lens is given by the power-law (PL) profile (we denote $\theta=|\vec\theta|$)
\be
\label{eq:PLmodel}\vec\alpha(\vec\theta)&=&\left(\frac{\theta}{\tilde\theta_{\rm E}}\right)^{1-\gamma_{\rm PL}}\vec\theta.
\ee
To this, we add some true physical values for $\kappa^{\rm s,ls,l}$, so altogether the imaging data satisfies Eqs.~(\ref{eq:lensweak}-\ref{tilde_alpha}). Note that because of weak lensing, the parameter $\tilde\theta_{\rm E}$ in Eq.~(\ref{eq:PLmodel}) is not equal to the Einstein angle, that we will denote by $\theta_{\rm E}$. 

The imaging part of the data can be summarised as a measurement of $\theta_{\rm E}$. We will simplify the discussion by assuming that also $\gamma_{\rm PL}$ is accurately determined.  
The effective modeling which transforms away the weak lensing terms would converge onto the model
\be\label{eq:alphamodeltoy} \vec\alpha^{\rm model}(\vec\theta)&=&\frac{(1-\kappa^{\rm ls})(1-\kappa^{\rm l})^{2-\gamma_{\rm PL}}}{1-\kappa^{\rm s}}\left(\frac{\theta}{\tilde\theta_{\rm E}}\right)^{1-\gamma_{\rm PL}}\vec\theta\no\\
&:=&\left(\frac{\theta}{\theta_{\rm E}}\right)^{1-\gamma_{\rm PL}}\vec\theta.\ee
The relation between the PL parameter $\tilde\theta_{\rm E}$ and the Einstein angle $\theta_{\rm E}$ is, therefore,
\be\label{eq:thetaEEmod}\theta_{\rm E} &=&\tilde\theta_{\rm E}\left[\frac{(1-\kappa^{\rm ls})(1-\kappa^{\rm l})^{2-\gamma_{\rm PL}}}{1-\kappa^{\rm s}}\right]^{\frac{1}{\gamma_{\rm PL}-1}} .
\ee

Turning to kinematics, the observable velocity dispersion for the PL profile is\footnote{We thank Daniel Johnson for pointing out the factor of $(1-\kappa^{\rm l})$ in the top line of Eq.~(\ref{eq:kins2}), which we erroneously missed in a previous version of this work.}
\be
\label{eq:kins2}
\sigma^2(\theta) &=& 2G \Sigma_{\rm crit} d_{\rm A}(0,z_{\rm l}) \frac{\sqrt{\pi}\Gamma\left(\frac{\gamma_{\rm PL}}{2}\right)}{\Gamma\left(\frac{\gamma_{\rm PL}-1}{2}\right)}\tilde\theta_E^{\gamma_{\rm PL}-1}((1-\kappa^{\rm l})\theta)^{2-\gamma_{\rm PL}} \no\\
&=& \frac{1-\kappa^{\rm s}}{1-\kappa^{\rm ls}} \frac{d_{\rm A}(0,z_{\rm s})}{d_{\rm A}(z_{\rm l},z_{\rm s})} J(\theta_{\rm E}, \gamma_{\rm PL}). 
\ee
In the second line we connect our result with Eq.~(8) of Ref.~\cite{Millon:2019slk} (see also~\cite{Birrer:2015fsm,Birrer:2018vtm}), defining $J$ as a cosmology-independent function that depends only on imaging observables.
For simplicity, we assume that the velocity dispersion is isotropic. 
The term $G\Sigma_{\rm crit} d_{\rm A}(0,z_{\rm l})=(1/4\pi)d_{\rm A}(0,z_{\rm s})/d_{\rm A}(z_{\rm l},z_{\rm s})$ is a function of the system redshifts and of cosmological parameters, but is independent of $H_0$ which cancels out in the ratio of angular diameter distances; for simplicity, we assume that it is known without error. Note that: (i) our derivation of Eq.~(\ref{eq:kins2})  accounts explicitly for the impact of weak lensing, so there are no hidden insertions of $\kappa^{\rm r}$ in the ratio $d_{\rm A}(0,z_{\rm s})/d_{\rm A}(z_{\rm l},z_{\rm s})$ which here simply expresses the ratio of the two usual redshift integrals defining $d_{\rm A}(z_{\rm o},z_{\rm e})$ in an unperturbed FRW cosmology, and (ii) from the first line in Eq.~(\ref{eq:kins2}), the kinematics measurement of $\sigma^2$ can be summarised as a measurement of $\tilde\theta_{\rm E}(1-\kappa^{\rm l})^{(2-\gamma_{\rm PL})/(\gamma_{\rm PL} -1)}$. 

Combining the kinematics data [$\tilde\theta_{\rm E}(1-\kappa^{\rm l})^{(2-\gamma_{\rm PL})/(\gamma_{\rm PL} -1)}$ via $\sigma^2$ in Eq.~(\ref{eq:kins2})] and the imaging data [$\theta_{\rm E}$ in Eq.~(\ref{eq:thetaEEmod})], one can obtain a measurement of the weak lensing factor, 
\be\frac{1-\kappa^{\rm ls}}{1-\kappa^{\rm s}}&=&\left[\frac{\theta_{\rm E}}{(1 - \kappa^{\rm l})^{\frac{2-\gamma_{\rm PL}}{\gamma_{\rm PL} -1}} \tilde\theta_{\rm E}}\right]^{\gamma_{\rm PL}-1} .\ee
This measurement is not equivalent to a measurement of the MSD factor 
\be \frac{1-\kappa^{\rm ls}}{(1-\kappa^{\rm s})(1-\kappa^{\rm l})}\nonumber\ee
that is needed in order to extract the truth value of $H_0$ from the effective model result $H_0^{\rm model}$ in Eq.~(\ref{eq:H0modcorrection}); specifically, even assuming that $\gamma_{\rm PL}$ is perfectly well known, the two weak lensing factors are offset by $1-\kappa^{\rm l}$ in the denominator.
 
Ref.~\cite{Millon:2019slk} presented a treatment of systematics in recent cosmographic analyses. 
There, the following expression was used to correct for the weak lensing MSD\footnote{See discussion around Eqs.(7-8) and Eq.~(16) in~\cite{Millon:2019slk}.}:
\be\label{eq:H0Millon19} H_0^{\rm inferred}&=&(1-\kappa^{\rm ext})H_0^{\rm model}.\ee
The terms $\kappa^{\rm ls,l}$ were effectively set to zero in the modeling, as they were ignored in both kinematics and imaging. From Eq.~(8) in Ref.~\cite{Millon:2019slk} and our Eq.~\eqref{eq:kins2} it follows that for a PL density profile, the term $\kappa^{\rm ext}$ should be identified with
\be\label{eq:kapextMillon19} 1-\kappa^{\rm ext}&:=&\frac{1-\kappa^{\rm s}}{1-\kappa^{\rm ls}}.\ee
This expression coincides with the discussion in Ref.~\cite{Birrer:2015fsm}, cited by~\cite{Millon:2019slk} for the treatment of kinematics, if we set $\kappa^{\rm ls}\to0$, in which case $\kappa^{\rm ext}\to\kappa^{\rm s}$.

Combining Eqs.~(\ref{eq:kapextMillon19}),~(\ref{eq:H0Millon19}), and~(\ref{eq:H0modcorrection}), we conclude that in Ref.~\cite{Millon:2019slk} the relation between the inferred value and the truth value of the Hubble parameter was biased by the following factor:
\be\label{eq:H0error}\frac{H_0^{\rm inferred}}{H_0}&=&\frac{1-\kappa^{\rm s}}{1-\kappa^{\rm ls}}\frac{1-\kappa^{\rm ls}}{(1-\kappa^{\rm s})(1-\kappa^{\rm l})}\no\\
&\approx&1+\kappa^{\rm l}.
\ee

We should note that although we considered $\sigma^2$ as an observable, in practice it is not directly measured. Various observational effects such as luminosity weighting, point spread function, etc., must be taken into account. Moreover, there are important theoretical uncertainties due to the velocity anisotropy, and also due to the actual lens halo density profile (even in the simple power law model considered above, unknown profile parameters include the slope $\gamma_{\rm PL}$), which must be marginalized over in the likelihood.

\section{On the use of ray tracing to resolve the MSD}\label{s:raymsd}
Another method to constrain external convergence, used in Refs.~\cite{Rusu:2019xrq,Birrer:2018vtm,DES:2019fny,Chen:2019ejq, Wong:2019kwg,Millon:2019slk}, is via ray-tracing in simulated data, calibrated system by system to the source density of the field containing the primary lens ~\cite{Keeton:1996tq,Holder:2002hq,Dalal:2004as,Momcheva:2005ex,Suyu:2009by,2017MNRAS.467.4220R,2018ApJ...867..107W,H0LiCOWX2019,2020MNRAS.498.1406T}. 

The correction for external convergence requires all of $\kappa^{\rm s,ls,l}$ to be extracted simultaneously, and applied to the cosmography analysis via Eq.~(\ref{eq:H0modcorrection}). However, Ref.~\cite{Rusu:2019xrq,Birrer:2018vtm,DES:2019fny,Chen:2019ejq, Wong:2019kwg,Millon:2019slk} only used ray tracing to derive the observer-source LOS term, $\kappa^{\rm s}$. This was identified in these analyses with the parameter $\kappa^{\rm ext}$, that was applied to correct for the effect in the determination of $H_0$ using Eq.~(\ref{eq:H0Millon19}), with $\kappa^{\rm ls,l}$ taken to vanish\footnote{As an aside, we note that the identification of $\kappa^{\rm ext}$ with $\kappa^{\rm s}$ extracted from ray tracing, and the alternative identification of $\kappa^{\rm ext}$ via kinematics as in Eq.~(\ref{eq:kapextMillon19}), are consistent for $\kappa^{\rm l}=\kappa^{\rm ls}=0$, but generally inconsistent otherwise.}
\footnote{The correct definition of $\kappa^{\rm ext}$ that incorporates all of $\kappa^{\rm s,ls,l}$ was explicitly written in Ref.~\cite{Birrer:2020tax} (we thank Simon Birrer for drawing our attention to this fact). However, also in~\cite{Birrer:2020tax}, when making contact with ray tracing priors it was assumed that $\kappa^{\rm ext}=\kappa^{\rm s}$; see Sec.~5.1 {\it there}.}.

Therefore we expect that in these analyses, the inferred value of $H_0$ (corrected by ray tracing for $\kappa^{\rm s}$) is still biased w.r.t. the truth value of $H_0$, by the amount:
\be \frac{H_0^{\rm inferred}}{H_0}&=&\frac{1-\kappa^{\rm ls}}{1-\kappa^{\rm l}}\;\approx\;1-\kappa^{\rm ls}+\kappa^{\rm l}.\ee

\begin{figure}
    \centering
    \includegraphics[scale=0.4]{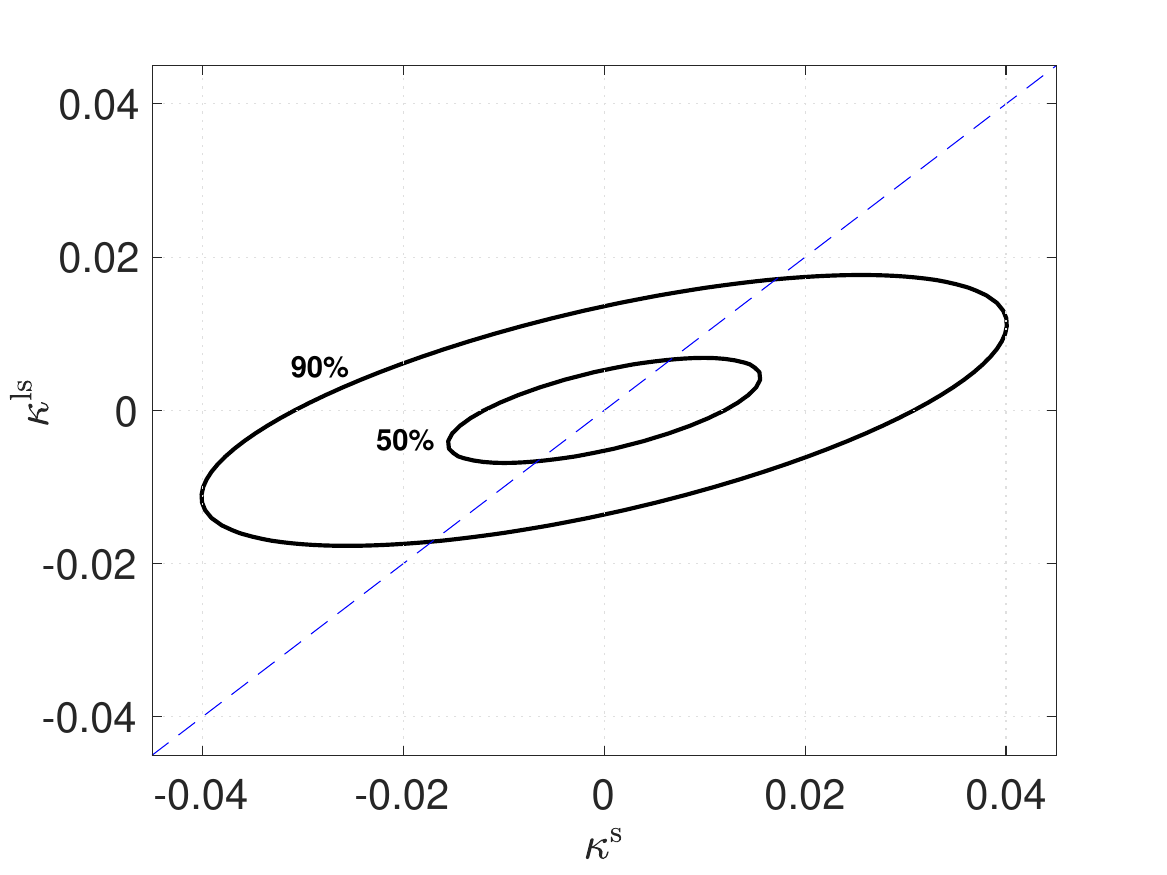}
    \includegraphics[scale=0.4]{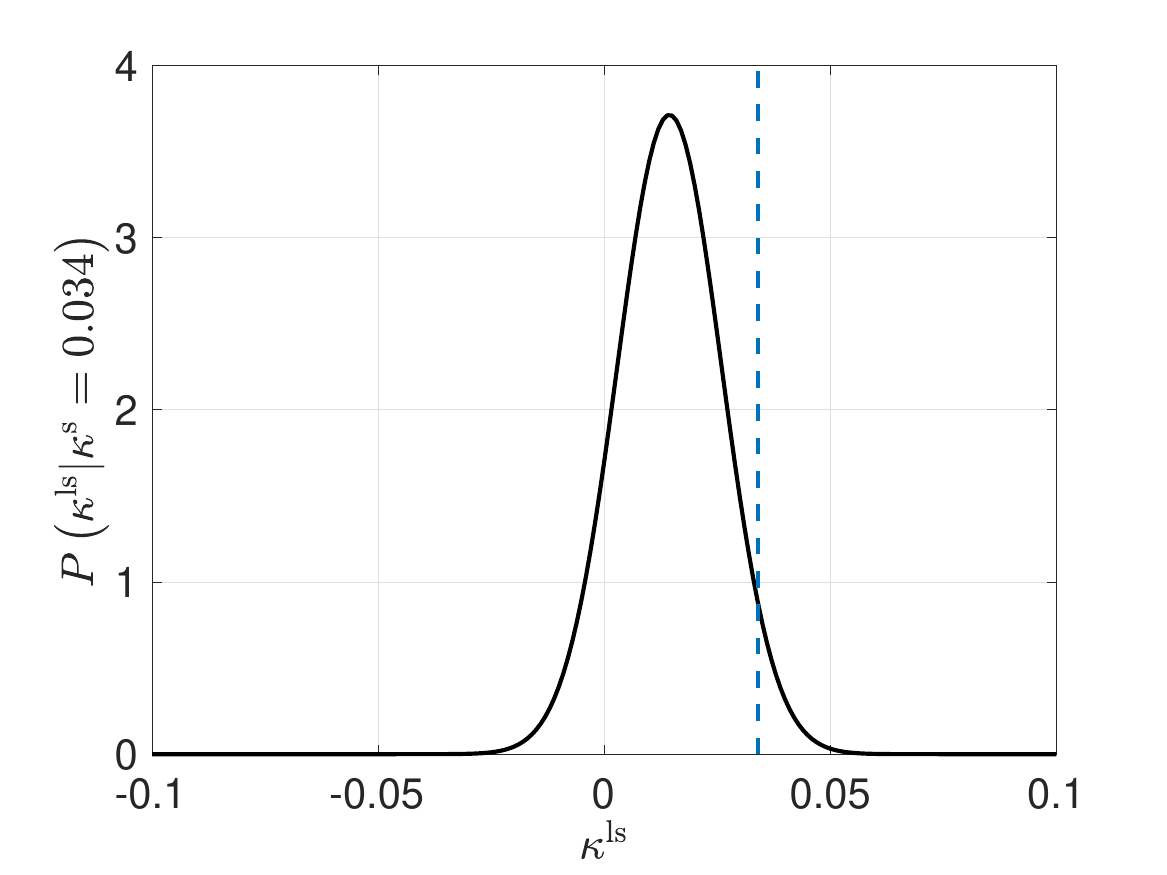}
    \caption{{\rm Top:} 50\% and 90\% joint probability quantiles for $\kappa^{\rm s}$ and $\kappa^{\rm ls}$, using the RMS values from Fig.~\ref{fig:kappa_ls_s_l} at $z_{\rm s}=2$. {\rm Bottom:} Conditional probability $P\left(\kappa^{\rm ls}|\kappa^{\rm s}=0.034\right)$. The reference value of $0.034$ is approximately equal to the RMS of $\kappa^{\rm s}$ at $z_{\rm s}=2$ in Fig.~\ref{fig:kappa_ls_s_l}. Note that this plot (like Fig.~\ref{fig:kappa_ls_s_l}) relies on a simplified model of the non-equal time matter power spectrum, and is not expected to be accurate beyond the $\mathcal{O}(1)$ level.}
    \label{fig:jointprob}
\end{figure}

We note that the $\kappa^{\rm s,ls,l}$ terms should be considered as separate (albeit statistically correlated) nuisance parameters in cosmography. To clarify this point, in Fig.~\ref{fig:jointprob} we show an estimate of the statistics of $\kappa^{\rm s}$ and $\kappa^{\rm ls}$ in a specific example (see, e.g.~\cite{Keeton:1996tq} and references in and of it for previous studies). For definiteness, for this example we use the results shown in Fig.~\ref{fig:kappa_ls_s_l} with a source redshift $z_{\rm s}=2$. The {\bf top} panel of Fig.~\ref{fig:jointprob} shows the 50\% and 90\% quantiles of the bivariate distribution of $\kappa^{\rm s},\kappa^{\rm ls}$, assuming Gaussian statistics. The {\bf bottom} panel shows the conditional probability distribution of $\kappa^{\rm ls}$ given a measured value of $\kappa^{\rm s}=0.034$ (corresponding to the RMS of $\kappa^{\rm s}$ in this example). We emphasize that our calculation here uses the analysis of App.~\ref{sec:kappa_variance}, and can only be used as a rough estimate of the statistics of the weak lensing terms. More accurate results probably require ray tracing simulations.

\section{Multiple sources and differential convergence} \label{sec:modeling_without_kappaext}

If multiple sources are available, then the MSD requires a simultaneous adjustment of the weak lensing variables for all sources. In particular, the $\kappa^{\rm s,ls}_i$ parameters for all sources $i$ must be adjusted together, following Eqs.~(\ref{eq:Mmsds},\ref{eq:Mmsdls}). 
Therefore, in the multi-source scenario, a certain combination of external convergence terms is measurable from the imaging data. 
A quick way to see what this measurable combination is, is by assuming that the analysis pipeline attempts to fit the two systems $i$ and $j$ separately and independently, omitting external convergence from the equations using Eq.~(\ref{eq:alphamodel}). The outcome of such a procedure are two independent fits for the  effective deflection angle, the results of which should be related by an over-all factor:
\be \label{alpha_ratio}
\frac{\left|\vec{\alpha}^{\rm model}_j(\vec\theta)\right|}{\left|\vec{\alpha}^{\rm model}_i(\vec\theta)\right|} &=&\frac{C_j}{C_i}\frac{1-\kappa^{\rm ls}_j}{1-\kappa^{\rm ls}_i}\frac{1-\kappa^{\rm s}_i}{1-\kappa^{\rm s}_j}\;\approx\;  \frac{C_j}{C_i}\left(1 + \delta\kappa^{\rm s}_{ji} - \delta\kappa^{\rm ls}_{ji}\right),\no\\
\ee
where $\delta\kappa^{\rm r}_{ji}:= \kappa^{\rm r}_{j} -\kappa^{\rm r}_{i}$. 
The left hand side of Eq.~\eqref{alpha_ratio} is measurable, and the $C_i$'s are known, so the combination $(\kappa^{\rm ls}_{j} -\kappa^{\rm ls}_{i})-(\kappa^{\rm s}_{j} -\kappa^{\rm s}_{i})$ is, in principle, measurable. Unfortunately, as this combination of terms is invariant under the MSD
, it cannot resolve the MSD impact on the $H_0$ inference.

\subsection{MSD-core (``internal convergence")}\label{ss:msdcore} 
Uncertainties in the intrinsic mass profile of the lens could pose a more serious problem to time-delay measurements of $H_0$, than that posed by weak external convergence. 
Specifically, an extended cored density component in lens galaxies would act similarly to external convergence~\cite{Schneider_2013, Blum:2020mgu}, but could, in principle, cause a much larger effect. 
This scenario could occur in some models of dark matter~\cite{Blum:2021oxj}.

To make the discussion concrete, consider the following change to the intrinsic physical surface mass density of the primary lens,
\be\label{eq:addcore}\Sigma(\vec x)&\to&\Sigma(\vec x)+\Sigma_{\rm c}(\vec x).\ee
We can think of the original density profile, $\Sigma(\vec x)$, as some steeply-falling mass distribution. It could come, for example, from the sum of a CDM Navarro-Frenk-White (NFW) profile, with $\Sigma_{\rm NFW}(r)\propto1/r^2$ at $r\gg R_{\rm S}$, where $R_{\rm S}$ is the NFW length scale parameter, and a stellar mass distribution $\Sigma_*(r)$ that falls even faster at large $r$. At smaller radii, near and around the projected Einstein radius of the lenses, lensing analyses often assume $\Sigma(r)\sim1/r$ (corresponding to 3D density scaling as $\rho\propto1/r^2$). 

In contrast, we will assume that the core component $\Sigma_{\rm c}(r)$ is nearly constant for $r$ near and below the projected Einstein radius. Note that by adding the core component in Eq.~\eqref{eq:addcore}, we are not eliminating the cusp of $\Sigma(r)$ at small $r$, but rather just adding to it a sub-dominant constant density term. At large radii, $r> R_{\rm c}$, the core component is assumed to decay, eventually joining or falling below the original $\Sigma(r)$. The lensing analyses constrain $R_{\rm c}$ to be larger than a few times the projected Einstein radius of the lens, with precise details of the transition depending on the precise implementation of the core profile~\cite{Blum:2021oxj,Birrer:2020tax}. For the ultralight DM cores considered in~\cite{Blum:2021oxj}, for example, lensing data demands that $R_{\rm c}$ should be larger than $\sim3$ times the projected Einstein radius of the lens. In what follows, for simplicity, we will assume that $R_{\rm c}$ is large enough so that we can neglect the finite radius corrections. Restoring these effects is straightforward, and not essential for our current analysis. 
Kinematics analyses~\cite{Cappellari2015} could also constrain a core feature, and may be able to provide an upper limit on $R_{\rm c}$, although the cusp+core composite model has not yet been included in existing studies. 

If $R_{\rm c}$ is large enough, then the core term in Eq.~\eqref{eq:addcore} is mathematically identical to a redefinition of the observer-source external convergence. Considering Eq.~\eqref{tilde_alpha}, we see that at the level of the modeling of imaging data, the core component is indistinguishable from the shift
\be M^{\rm s}_i&\to&M^{\rm s}_i+(\mathbb{I} - M^{\rm ls}_{i} )(\mathbb{I} - M^{\rm l})C_i\kappa_{{\rm c}1},\ee
where
\be\kappa_{{\rm c}1}&=&\frac{\Sigma_{\rm c}}{\Sigma_{\rm crit}(z_{\rm l},z_1)}.\ee
We will think of the internal core convergence $\kappa_{\rm c1}$ as a small parameter, albeit potentially somewhat larger than cosmological weak external convergence terms. We have in our mind the lensing contribution to the $H_0$ tension~\cite{Verde:2019ivm,DiValentino:2021izs}, that could be resolved by $\kappa_{\rm c1}\approx0.1$~\cite{Blum:2020mgu,Blum:2021oxj}.

Expressed in terms of convergence and shear parameters, at leading order in weak lensing terms, we have
\be\label{eq:kappacore}\kappa^{\rm s}_i&\to&\kappa^{\rm s}_i+C_i\kappa_{{\rm c}1}(1-\kappa^{\rm ls}_i-\kappa^{\rm l}),\\
\label{eq:gammacore}\Gamma^{\rm s}_i&\to&\Gamma^{\rm s}_i-C_i\kappa_{{\rm c}1}\left(\Gamma^{\rm ls}_i+\Gamma^{\rm l}\right).\ee
With this understanding one can see that imaging data alone cannot directly separate a core component from weak lensing. One must resort to kinematics analyses, or to theoretical considerations that could limit the plausible weak lensing effect. We focus on the latter.

It is worthwhile to highlight a key difference between convergence and shear. Under Eqs.~(\ref{eq:Mmsds}-\ref{eq:Mmsdl}), which deal purely with the modeling of external weak lensing, shear is adjusted multiplicatively, while convergence receives an additive correction. This feature is modified in Eqs.~(\ref{eq:kappacore}-\ref{eq:gammacore}), but a key part of it remains manifest: the addition of the core adjusts $\Gamma^{\rm s}$ via an additive term, however that additive term is itself proportional to the shear terms $\Gamma^{\rm ls}+\Gamma^{\rm l}$. As a result, even if $\kappa_{\rm c1}$ is somewhat larger than typical weak lensing effects (e.g. $\kappa_{\rm c1}\approx0.1$), this still only amounts to a relative correction of $\sim$10\% in $\Gamma^{\rm s}$. Constraining such a small effect observationally or theoretically would be  challenging. This point is important because certain combinations of weak lensing shear terms can, in principle, be measured directly from imaging data~\cite{Birrer:2016xku,Fleury:2021tke}. 
For convergence, Eq.~\eqref{eq:kappacore} suggests a potentially large additive readjustment of $\kappa^{\rm s}$, if $\kappa_{\rm c1}$ is larger than typical weak lensing effects. However, if only one source is available ($i=1$), then it could be difficult for imaging data alone to constrain $\kappa_{\rm c1}$. 

If more than one source is available, then we have seen in Sec.~\ref{sec:modeling_without_kappaext} that a certain combination of differential convergence terms is measurable from the imaging data. Inserting Eq.~\eqref{eq:kappacore} into Eq.~\eqref{alpha_ratio}, and neglecting the small correction factor $1-\kappa^{\rm ls}_i-\kappa^{\rm l}\approx1$ in Eq.~\eqref{eq:kappacore}, we see that the following ratio of deflection angles can be measured:
\be \label{alpha_ratiocore}
\frac{C_i}{C_j}\frac{\left|\vec{\alpha}^{\rm model}_j\right|}{\left|\vec{\alpha}^{\rm model}_i\right|} &\approx&  1 + (\kappa^{\rm s}_j-\kappa^{\rm s}_i) - (\kappa^{\rm ls}_j-\kappa^{\rm ls}_i)+\kappa_{\rm c1}\left(C_j-C_i\right).\no\\&&
\ee

At a first glance in Eq.~\eqref{alpha_ratiocore}, one could hope that multiple source systems could resolve the core-MSD ambiguity, because the last term on the right hand side contains the large additive term $\propto\kappa_{\rm c1}$. However, a second glance reveals a setback: in Eq.~\eqref{alpha_ratiocore}, $\kappa_{\rm c1}$ appears multiplied by the factor $C_j-C_i$, proportional to the relative difference of angular diameter distance combinations of the two sources (see Eq.~\eqref{eq:aia1}). Unfortunately, the angular diameter distance is a non-monotonous function of redshift; moreover, the sources of typical strong lensing systems are often located between $z\sim1$ and $z\sim2.5$, that is, around the shallow maximum of $d_{\rm A}(0,z)$. As a result, in many systems of interest, the difference $C_j-C_i\sim\mathcal{O}(0.1)$ is much smaller than unity. This ``dilutes" the efficiency at which imaging data in multiple source systems could constrain the internal MSD.

\begin{figure}
    \centering
    \includegraphics[scale=0.4]{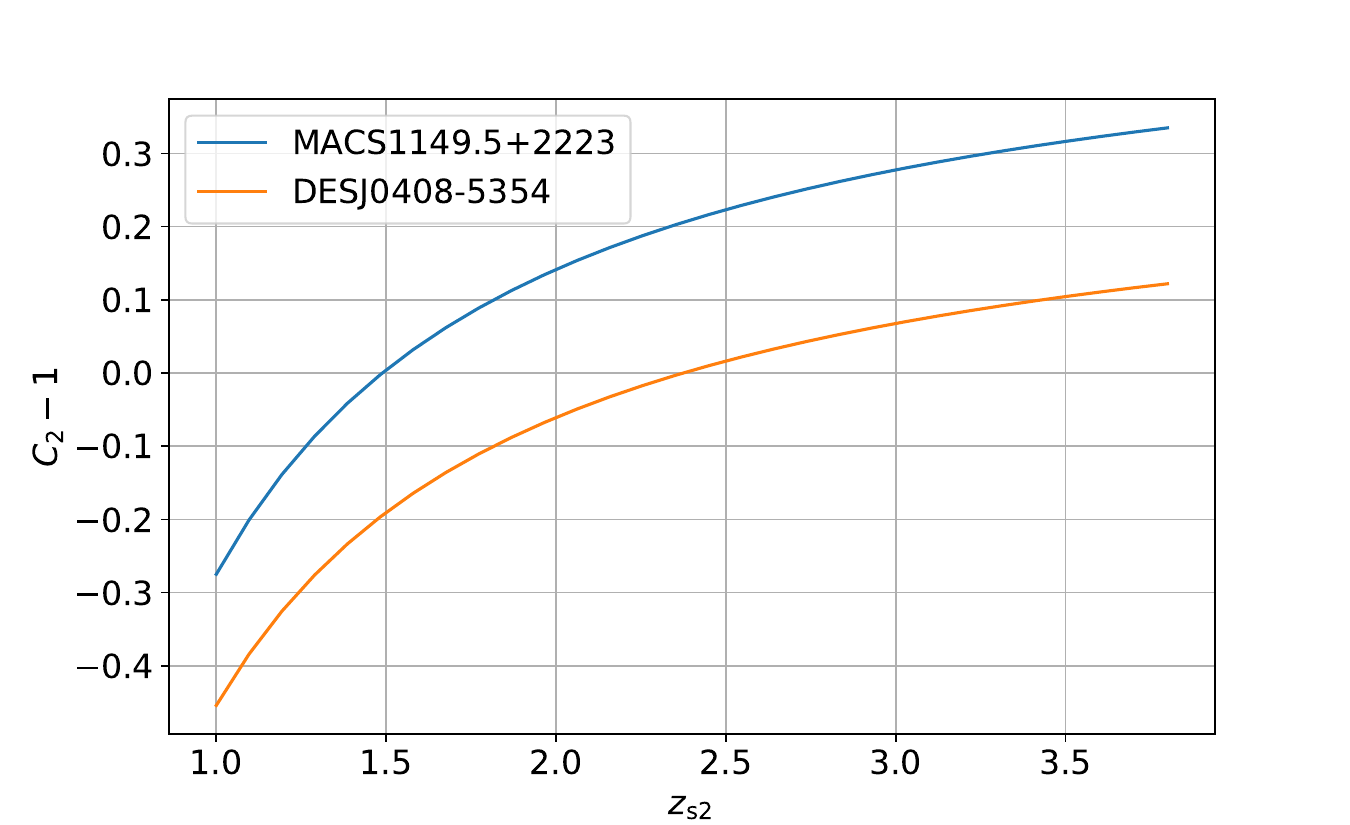}
    \caption{The angular diameter distance combination $C_2$, appearing in lensing analyses with multiple sources. Code: \href{https://github.com/lucateo/Comments_MSD/blob/main/Notebooks/delta_kappa_nonlinear.ipynb}{\faGithub}.
    }
    \label{fig:C_plot}
\end{figure}

Fig.~\ref{fig:C_plot} illustrates our point. We show two examples of the curve $C_2-1$. The blue line is inspired by the multiple source system of the cluster lens MACS1149.5+2223~\cite{Grillo:2018ume,Grillo:2020yvj}. The primary lens (cluster) redshift is $z_{\rm l}\approx0.5$. Time-delays are measured for a type Ia supernova (source 1) at $z_1\approx1.5$. Fig.~\ref{fig:C_plot} shows $C_2-1$ as function of a second source redshift $z_2$. (Actual additional sources of this system are distributed between $z_2\sim1.2$ and $z_2\sim3.7$.) The orange line is inspired by the galaxy lensing system DES J0408-5354~\cite{DES:2019fny}, $z_{\rm l}\approx0.6$, with time-delays measured to a quasar at $z_1\approx2.3$. 

Because $C_j-C_i$ is a small number, the $\kappa_{\rm c1}$ term in Eq.~\eqref{alpha_ratiocore} could be diluted down to the natural scale of weak cosmological convergence. To detect (or constrain) an internal core, it therefore becomes crucial to estimate the magnitude of weak differential convergence. We consider this problem in App.~\ref{sec:kappa_variance}, and comment on examples in Sec.~\ref{sec:real_systems}.

Before we move on, let us make a rough assessment of the precision by which the left hand side of \eqref{alpha_ratiocore} can actually be measured. Note that most of the information in the lensing data comes from the angular range $\theta\sim\theta_{\rm E}$, where for simplicity of this estimate we can consider spherically symmetric systems and drop the vector notation on $\theta$. Using the fact that $\alpha(\theta_{\rm E})=\theta_{\rm E}$, the relative uncertainty by which $|\vec\alpha_j/\vec\alpha_i|$ can be measured is of similar size as the quoted precision on the ratio of Einstein angles, $|\theta_{{\rm E}j}/\theta_{{\rm E}i}|$. For typical TDCOSMO systems, this precision is at the level of $\sim1\%$. Of course, this quoted precision corresponds to the main source considered by the analysis (usually, the source for which time-delays are measured). What we actually need is the differential convergence, and the precision on that would be dominated, given two sources $i=1,2$, by the source for which the precision on $\theta_{{\rm E},i}$ is poorest.

\subsection{Examples of multiple source systems} \label{sec:real_systems}
A key point of our analysis is that the availability of multiple sources in a lensing system can only resolve the core-MSD degeneracy to the extent, that the core-induced term, $\kappa_{\rm c1}(C_i-C_j)$, is significantly larger than the natural expectation for the weak cosmological differential convergence term, $\delta\kappa^{\rm s}_{ij}-\delta\kappa^{\rm ls}_{ij}$, in Eq.~\eqref{alpha_ratiocore}. Having armed ourselves, in App.~\ref{sec:kappa_variance} and App.~\ref{sec:compute_kappa}, with an estimate for the external convergence, we now explore two multi-source systems from the literature.

\begin{figure}
    \centering
    \includegraphics[scale=0.4]{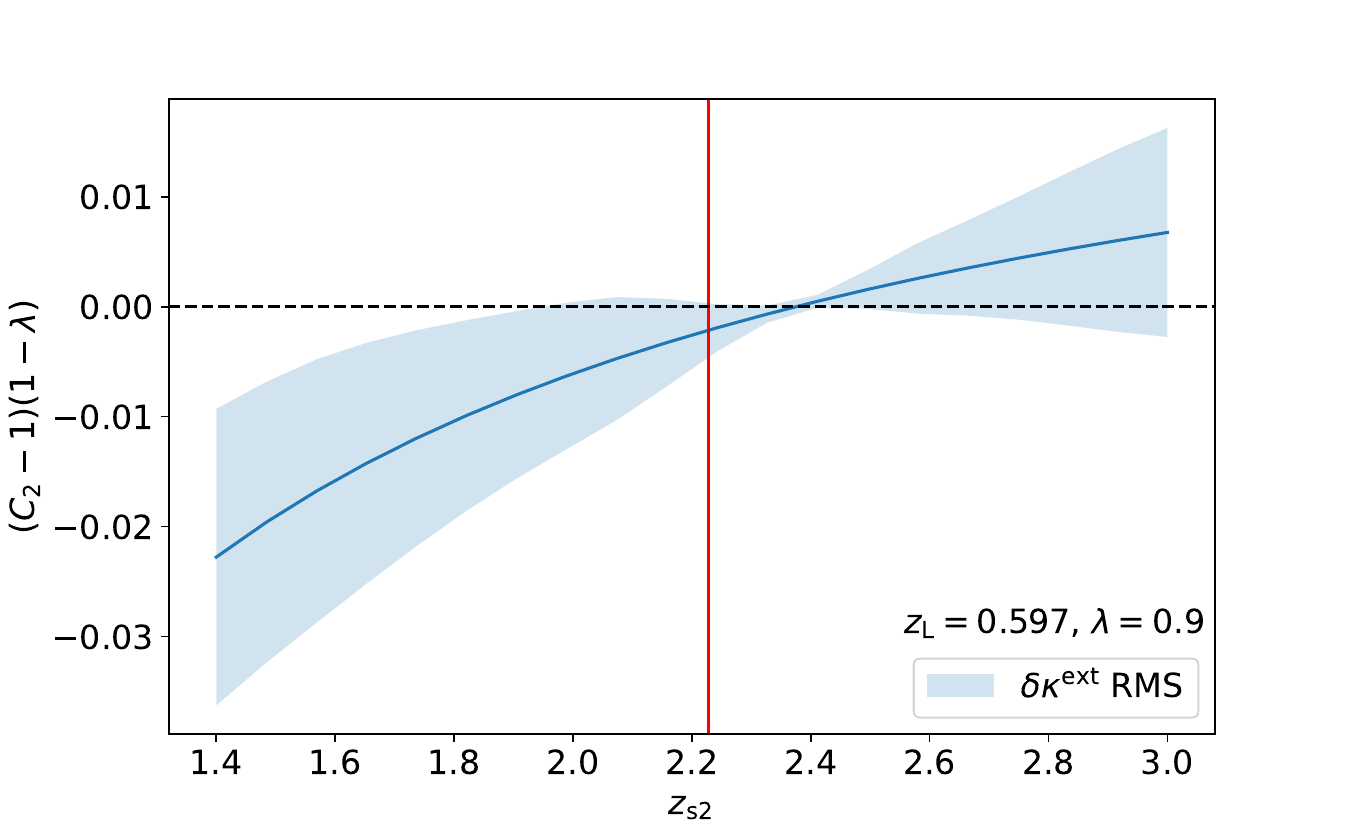}
    \includegraphics[scale=0.4]{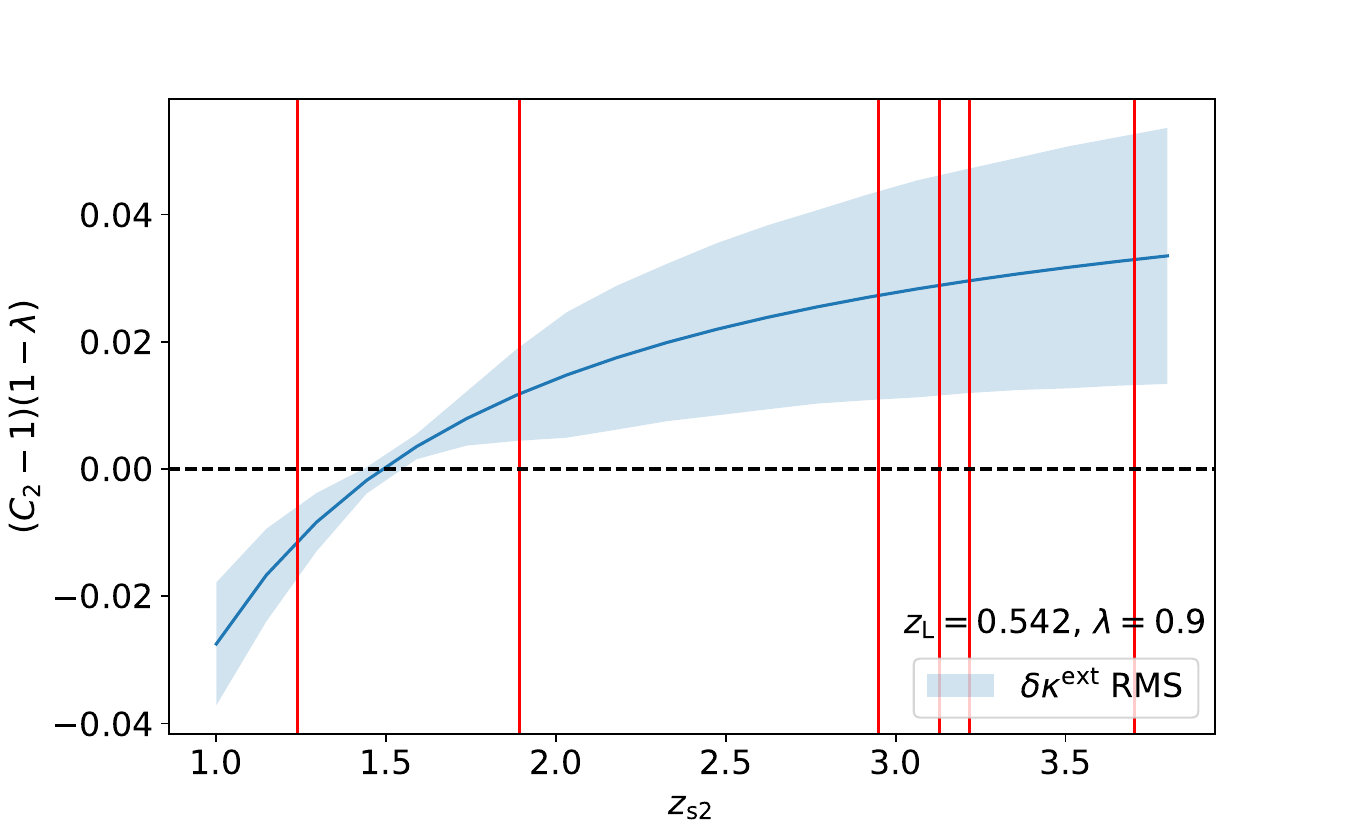}
    \caption{The angular diameter function $C_2-1$, weighted by a factor $\kappa_{\rm c1}\equiv1-\lambda=0.1$ (chosen to mimic a resolution of the lensing $H_0$ tension), compared with the cosmological RMS weak differential convergence $\delta\kappa$. {\bf Top:} redshift parameters chosen to resemble the TDCOSMO system DESJ0408-5354 \cite{DES:2019fny}. {\bf Bottom:} parameters chosen to resemble the MACS J1149.5+2223 cluster system  \cite{Grillo:2018ume,Grillo:2020yvj}. In both panels, the function $C_i-1$ vanishes at the redshift of the primary source (the source to which time-delays are measured). Vertical red lines mark the redshifts of secondary sources, that one could try to use to resolve the core-MSD. Code: \href{https://github.com/lucateo/Comments_MSD/blob/main/Notebooks/delta_kappa_nonlinear.ipynb}{\faGithub}.}
    \label{fig:kappa_variance}
\end{figure}

\subsubsection{DESJ0408-5354}
As noted earlier, this galaxy lensing system has a primary lens at $z_{\rm l}\approx0.6$, and main source (lensed quasar-host galaxy) at $z_1\approx2.3$, and a secondary source at $z_2\approx2.2$.
A TDCOSMO analysis of this system, fitting an elliptic power-law density model for the lens (without allowing for a core component), inferred a value of $H_0$ which was $\approx11$\% higher than the CMB/LSS result~\cite{DES:2019fny}. Thus, a core component at $\kappa_{\rm c1}\approx0.1$ could completely resolve the lensing $H_0$ tension for this system~\cite{Blum:2020mgu,Blum:2021oxj}. 

The question arises, whether the presence of the second lensed source for this system could resolve the core-MSD associated with $\kappa_{\rm c1}$. To address this question, in the {\bf top} panel of Fig.~\ref{fig:kappa_variance} we plot (blue line) the function $C_2-1$ for this system, weighted by the factor $\kappa_{\rm c1}\equiv1-\lambda=0.1$. To demonstrate the confusion with weak external convergence, following Eq.~\eqref{alpha_ratiocore} we superimpose a band with width chosen as the RMS value of $\delta\kappa^{\rm s}_{12}-\delta\kappa^{\rm ls}_{12}$ for the system. The red vertical line marks the redshift of the actual secondary source.

We conclude that multi-source imaging data for DESJ0408-5354~\cite{DES:2019fny} is unlikely to help in constraining the core-MSD proposal sufficiently to solve the lensing $H_0$ tension. 

\subsubsection{MACS J1149.5+2223}
As noted earlier, the lens in MACS J1149.5+2223~\cite{Treu:2015poa,Grillo:2018ume,Grillo:2020yvj} is a galaxy cluster at $z_{\rm l}\approx0.54$. The main source is a type-Ia supernova at $z_1\approx1.5$. Six additional multiply-imaged sources are distributed in redshift in the range $z_i\approx1.2$ to $z_i\approx3.7$. 

In the {\bf bottom} panel of Fig.~\ref{fig:kappa_variance} we show that for the secondary sources in MACS  J1149.5+2223~\cite{Grillo:2018ume,Grillo:2020yvj}, weak differential convergence should significantly (although, perhaps, not entirely) mask the presence of an internal MSD. 
We thus expect that adding differential convergence as nuisance parameters for the secondary sources (that is, the sources additional to the SNIa host, to which time delays were specified in the mock of~\cite{Grillo:2020yvj}) would significantly increase the uncertainty on the impact of the MSD as compared to the preliminary results in the appendix of~\cite{Grillo:2020yvj}.

\section{Summary}\label{s:sum}
In the effort to determine the Hubble parameter $H_0$ using strong lensing time delays, a key challenge is the mass sheet degeneracy (MSD).  
The MSD can be naturally associated with two physical phenomena: cosmological weak lensing (``external convergence" or ``external MSD"); and the possibility of a core component in the lens object (``internal MSD"). 
Well known methods to alleviate the MSD are: (i) the combination of imaging data with stellar kinematics, (ii) the use of ray tracing simulations to obtain an observationally-informed theoretical prior on external weak lensing, and (iii) the study of systems containing more than one strongly-lensed source. 

In this paper we discussed some issues related to the MSD. In Sec.~\ref{s:kinmsd}, regarding the use of kinematics, we noted that the relation between kinematics constraints and imaging data involves a combination of weak lensing terms that includes all of the observer-source, observer-lens, and source-lens segments of the line of sight (LOS). Neglecting the source-lens and observer-lens convergence terms -- a common practice in current analyses -- could lead to a bias of the order of a few percent in the inference of $H_0$ from time delays. It is possible to account for the effect by adding the observer-lens term as nuisance parameter in the combined imaging+kinematics likelihood.

In Sec.~\ref{s:raymsd} we noted that the neglect of the source-lens and observer-lens LOS contributions also affects ray tracing methods. Here too, omitting some of the LOS terms should bias the $H_0$ inference. It should be possible to extract priors for all of the LOS terms, and not only the observer-source one, from ray tracing. 

As we review in Sec.~\ref{s:msdweak}, the MSD is not broken by the availability of multiple sources in the imaging analyses. 
In Sec.~\ref{sec:modeling_without_kappaext} we considered what multiple sources do allow one to measure, which is differential convergence between different sources. Interestingly, weak differential external convergence complicates attempts to resolve the internal MSD, even if the internal core effect is parameterically larger than the weak lensing terms. The problem is that multiple sources are only useful against the internal MSD to the extent that they come with significantly different angular diameter distances; in practice, however, the angular diameter distances in typical multi-source systems used in cosmography are similar to the 10\% level.

In App.~\ref{sec:kappa_variance} we described a non-perturbative calculation of cosmological external convergence, that allows us to provide rough estimate of the expected size of the effect, as well as estimates of statistical correlations between different convergence terms. Our calculation suggests (what we think is) a natural approximate way to account for non-linear matter power spectra entering in correlation functions at different values of the cosmic time variable.

\acknowledgments
We are grateful to Fred Courbin and especially Simon Birrer for comments on the manuscript, including spotting a mistake in our preliminary draft. We are also grateful to Daniel Johnson for spotting a mistake in a previous version of this work.
The work of KB was supported by grant 1784/20 from the Israel Science Foundation. The work of YS was supported by grants from the NSF-BSF (No.~2018683), the ISF (No.~482/20), the BSF (No.~2020300) and by the Azrieli foundation. 
The work was supported by the International Helmholtz-Weizmann Research School for Multimessenger Astronomy, largely funded through the Initiative and Networking Fund of the Helmholtz Association. 
This work made use of the following public software packages: CAMB \cite{Lewis:1999bs,Lewis:2002ah}, pyfftlog (based on Ref.~\cite{Hamilton:1999uv}). 

\appendix

\section{The lens equation with weak lensing} \label{sec:deriveLSS}
In this appendix we review the derivation of the weak lensing effects in the lens equation. These results are known~\cite{Bartelmann:1999yn,1991ApJ...380....1M,Kaiser:1992ps,Bar-Kana:1995qyu,Keeton:1996tq,McCully:2013fga,Schneider:2014vka,Fleury:2021tke}, and we include them here for completeness of the main text. Let us suppose that we have a strong deflector located at a comoving distance $ \eta_{\rm l} $. We can split the gravitational potential as 
\begin{equation}
\Phi(\vec\beta(\eta), \eta) = \tilde{\Phi}(\vec\beta(\eta_{\rm l}), \eta_{\rm l}) \delta(\eta- \eta_{\rm l}) + \Phi_{\mathrm{t}} (\vec\beta(\eta), \eta)  ,
\end{equation}
where $ \Phi_{\mathrm{t}} (\beta(\eta), \eta) $ is the weak gravitational potential associated to weak lensing effects, and $\tilde\Phi$ is the gravitational potential of the main deflector. We can implement the tidal approximation on $ \Phi_{\mathrm{t}} $ by setting
\be\label{Phi_tidal}
\Phi_{\rm t}(\vec\beta(\eta), \eta) &\approx& \Phi_{\rm t}(0, \eta) + \beta_i \partial_i \Phi_{\mathrm{t}} (0, \eta)   .
\ee

The lens equation may be written as~\cite{Bartelmann:2010fz}
\be
\beta_i(\eta) &=& \theta_i - 2 \int_{0}^{\eta} \dd{\eta'} \frac{\eta - \eta' }{ \eta \eta'} \partial_i \Phi (\vec{\beta}(\eta'), \eta').
\ee
Within the tidal approximation, Eq.~\eqref{Phi_tidal}, we can write
\be\label{beta_wl}
\beta_i(\eta_{\rm l}) &=& \theta_i - 2 \int_{0}^{\eta_{\rm l}} \dd{\eta'} \frac{\eta_{\rm l} - \eta' }{ \eta_{\rm l} \eta'} \partial_i \Phi_{\rm t} (0, \eta') \no\\
&-& 2 \int_{0}^{\eta_{\rm l}} \dd{\eta'} \frac{\eta_{\rm l} - \eta' }{ \eta_{\rm l}\eta'} \partial_i \partial_j \Phi_{\rm t} (0, \eta') \beta_j(\eta')  .
\ee
The second term on the RHS of Eq.~(\ref{beta_wl}) is an unobservable overall shift of the deflection angle (independent of $ \vec\theta $), which can be reabsorbed in the source coordinates.
Defining
\be\label{M_ij}
M_{ij}(\eta_1, \eta_2) &:=& 2  \int_{\eta_1}^{\eta_2} \dd{\eta'} \frac{(\eta_2 - \eta') (\eta' - \eta_1) }{ (\eta_2 - \eta_1)\eta^{'2}} \partial_i \partial_j \Phi_{\rm t} (0, \eta'),\no\\&&
\ee
we expect $ M_{ij} $ terms to be small as long as we are dealing with weak fields and maintain only terms at first order in these quantities. In particular, for $ \eta < \eta_{\rm l} $, substituting
\be
\beta_i(\eta) &=& \theta_i - 2 \int_{0}^{\eta} \dd{\eta'} \frac{\eta - \eta' }{ \eta \eta'} \partial_i \partial_j \Phi_{\rm t} (0, \eta')\beta_j (\eta')   \;\;\;\;
\ee
in Eq.~\eqref{beta_wl}, we obtain
\be
\vec\beta(\eta_{\rm l}) &=&(\mathbb{I} - M(\eta_{\rm l}, 0)) \vec\theta  .
\ee
For $ \eta > \eta_{\rm l} $, the situation changes due to the presence of the strong deflector. Considering the full $ \Phi $ from Eq.~\eqref{Phi_tidal}, avoiding the tidal approximation for the strong deflector (but using the thin lens approximation, encoded in the Dirac delta), we have, with $\eta_{\rm s}$ as the comoving distance of the source, 
\begin{align} \label{beta}
\begin{aligned}
\beta_i&(\eta_{\rm s}) = \theta_i - 2 \int_{0}^{\eta_{\rm s}} \dd{\eta'} \frac{\eta_{\rm s} - \eta' }{ \eta_{\rm s}\eta'} \partial_i \partial_j \Phi_{\rm t} (0, \eta') \beta_j(\eta') \\
- &\underbrace{2 \frac{\eta_{\rm s} - \eta_{\rm l} }{ \eta_{\rm s} \eta_{\rm l}} \partial_i \tilde{\Phi}(\vec\beta(\eta_{\rm l}))}_{= \alpha_i(\vec\beta(\eta_{\rm l}))} = (\delta_{ij} -  M_{ij}(\eta_{\rm l}, 0)) \theta_j - \alpha_i(\vec\beta(\eta_{\rm l})) \\
+& 2\int_{\eta_{\rm l}}^{\eta_{\rm s}} \dd{\eta'} \frac{\eta_{\rm s} - \eta' }{ \eta_{\rm s}\eta'} \partial_i \partial_j \Phi_{\rm t} (0, \eta') \qty[ 2 \frac{\eta' - \eta_{\rm l} }{ \eta_{\rm l}\eta'} \partial_i \tilde{\Phi}(\vec\beta(\eta_{\rm l})) ] ,
\end{aligned}
\end{align}
where on the last step we substituted $ \vec\beta(\eta') $ inside the integral with the term 
\be
\beta_i(\eta') &=& \begin{cases}
&\theta_i  \text{ for } \eta\leq \eta_{\rm l}  , \\
&\displaystyle\theta_i - 2 \frac{\eta' - \eta_{\rm l} }{ \eta_{\rm l}\eta'} \partial_i \tilde{\Phi}(\vec\beta(\eta_{\rm l})) \text{ for } \eta > \eta_{\rm l}  . 
\end{cases}\;\;\;\;
\ee
We can rewrite the term in square brackets in Eq.~\eqref{beta} as
\be
2 \frac{\eta' - \eta_{\rm l} }{\eta_{\rm l}\eta'} \partial_i \tilde{\Phi}(\vec\beta(\eta_{\rm l})) &=& \frac{(\eta' - \eta_{\rm l}) \eta_{\rm s}}{\eta' (\eta_{\rm s} - \eta_{\rm l}) } \alpha_i(\vec\beta(\eta_{\rm l})) ,\;\;\;
\ee 
finally arriving at Eq.~\eqref{weakLens_original} in the form
\be\label{lens_eq weak lensing}
\vec\beta(\eta_{\rm s}) &=& (\mathbb{I} - M(\eta_{\rm s}, 0)) \vec\theta \no\\
&-& (\mathbb{I} - M(\eta_{\rm s}, \eta_{\rm l}))\vec\alpha\qty((\mathbb{I} - M(\eta_{\rm l}, 0)) \vec\theta) .\;\;
\ee

The time delay between image solutions of Eq.~\eqref{lens_eq weak lensing} can be computed by exploiting the Fermat principle~\cite{Schneider:1992,Schneider:1997bq,Schneider:2014vka}. First, note that we can write the potential part of the time delay due to the main deflector, $t_{\rm pot}$, as
\be \label{time_delay_prefactor}
t_{\rm pot} &=& - D_{\rm dt} \psi ((\mathbb{I} - M^{\rm l})\vec{\theta} )  .
\ee
The Fermat principle states that, up to an affine transformation, the lens equation can be obtained by taking the gradient $\grad_{\vec{\theta}}$ of the time delay function $t(\vec{\theta}, \vec{\beta})$ and setting it to zero. Eq.~\eqref{time_delay_prefactor} can then be used to understand what is the correct prefactor (the affine parameter) entering the time delay function. We see that from the function
\be \label{t_func}
t(\vec{\theta}, \vec{\beta}) &=& D_{\rm dt} \Big(\frac{1}{2}\vec\theta^T\left(\mathbb{I}-M^{\rm s}-M^{\rm l}+M^{\rm ls}\right)\vec\theta \no\\
&-&\vec\beta^T\left(\mathbb{I}-M^{\rm l}+M^{\rm ls}\right)\vec\theta-\psi((\mathbb{I}-M^{\rm l})\vec\theta) \Big) ,\;\;\;\;\;\;\;
\ee
one indeed recovers Eq.~\eqref{lens_eq weak lensing} using $\grad_{\vec{\theta}} t(\vec{\theta}, \vec{\beta}) = 0$, recalling the definition $\grad_{\vec\xi}{\psi (\vec\xi)} = \vec{\alpha}(\vec\xi)$. Notice that Eq.~\eqref{t_func} has the correct prefactor, Eq.~\eqref{time_delay_prefactor}, in front of $\psi ((\mathbb{I} - M^{\rm l})\vec{\theta} )$. Finally, Eq.~\eqref{eq:DtAB} is recovered via $\Delta t_{AB} = t(\vec{\theta}_A, \vec{\beta}) - t(\vec{\theta}_B, \vec{\beta})$.

\subsection{Multi-plane lens equation.}\label{ssa:multiplane}
In our discussion, we did not take into account the possibility that nearer sources could act as additional lens planes for further sources~\cite{McCully:2013fga,Schneider:2014vka}. It should be clear that adding this effect into the modeling increases the complexity and also adds more possible layers of degeneracy, beyond and on top of the weak lensing MSD we emphasized in our analysis. Here we briefly explain how the effect can be embedded into our notation.

Adjusting our notation to that in  Ref.~\cite{Schneider:2014vka}, we label with the index $i=0$ the primary lens plane 
and with index $i>0$ the source planes, with $i>j$ implying that source $i$ has bigger redshift than source $j$. $\hat\alpha_i$ is now the deflection angle due to lens/source $i$, which relates with the usual quantity used in lens equations, $\vec\alpha_i$, with
\begin{equation} \label{new_alpha}
\vec\alpha_i = \frac{d_{\rm A}(z_i,z_{i+1})}{d_{\rm A}(0,z_{i+1})} \hat\alpha_i \ .
\end{equation}
With this, we can write the multi-plane lens equation as
\begin{align}
\begin{aligned}
\vec{\beta}_i = (\mathbb{I} - M(\eta_i, 0)) \vec\theta - \sum_{j=0}^{i-1} (\mathbb{I} - M(\eta_i, \eta_j) )C_{ji}\vec{\alpha}_j(\vec{\beta}_j)  ,  
\end{aligned}
\end{align}
where $M(\eta_i, \eta_j)$ is defined in Eq.~\eqref{M_ij} and
where $C_{ji}$ is the generalization of the factor in Eq.~\eqref{eq:aia1} coming from the definition Eq.~\eqref{new_alpha},
\begin{equation}
C_{ji} := \frac{d_{\rm A}( z_j, z_{i})\,d_{\rm A}(0, z_{j+1})}{d_{\rm A}(0, z_{i})\,d_{\rm A}(z_{j},z_{j+1})} .
\end{equation}

To incorporate these results into our discussion in the main text, one only needs to add to the MSD of  Eqs.~(\ref{msd_transform1}, \ref{msd_transform2}) the further requirement (for $i>1$)
\begin{align}
\begin{aligned}
\sum_{j=1}^{i-1} (\mathbb{I} & - M^{\lambda}(\eta_i, \eta_j)) C_{ji}\vec{\alpha}^{\lambda}_j(\vec{\beta}^{\lambda}_j )  \\ 
&=  \lambda\sum_{j=1}^{i-1}(\mathbb{I} - M(\eta_i, \eta_j)) C_{ji}\vec{\alpha}_j(\vec{\beta}^{\lambda}_j/\lambda) .
\end{aligned}    
\end{align}
This is a stretch of the argument in $\vec{\alpha}_j$ along with an over-all rescaling of $\vec{\alpha}_j$ and/or  $M(\eta_i, \eta_j)$.

\section{Cosmological external convergence} \label{sec:kappa_variance}
The cosmological external convergence between comoving distance $ \eta_{1} $ and $\eta_2 > \eta_1$ in the direction $\hat n$ on the sky can be written as (see Ref~\cite{Bartelmann:1999yn} and Eq.~\eqref{M_ij}):
\begin{align}
&\kappa(\eta_2, \eta_1; \hat{n})=\frac{3H^2_0\Omega_{\rm m}}{2}\int \dd{\eta} q_{21}(\eta) \delta(\hat n,\eta) , \\
&q_{ij}(\eta) := \Theta ( \eta - \eta_j) \Theta ( \eta_i - \eta)\frac{(\eta_{i}-\eta)(\eta - \eta_{j})}{\eta_{i} - \eta_{j}}(1 + z(\eta))   ,
\end{align}
where $\delta(\hat n,\eta)$ is the matter overdensity at $\vec x=\eta\hat n$,
\be\eta(z)&=&\frac{1}{H_0}\int_0^z\frac{\dd{z'}}{\sqrt{\Omega_\Lambda+\Omega_{\rm m}(1+z')^3}}\ee
is our comoving distance to the shell at $z$, and we have neglected 3-curvature and radiation in the cosmic energy budget.

To calculate RMS differential convergence, $ \sqrt{\expval{\delta\kappa_i^2}}$, we need to evaluate mixed correlation terms of the form
\begin{align}
\begin{aligned}
&\expval{\kappa(\eta_i, \eta_j; \hat{n}) \kappa(\eta_l, \eta_m; \hat{n}')} =\frac{9 H_0^4 \Omega^2_{\text{m},0}}{4} \int \dd{\eta} \int \dd{\eta'} \\
&\times q_{ij}(\eta) q_{lm}(\eta')\expval{\delta(\hat n,\eta)\delta(\hat n',\eta') }.
\end{aligned}
\end{align} 
Passing to Fourier space, and using the power spectrum
\begin{equation}
\expval{\delta(\vec n,\eta)\delta(\vec n',\eta')} = (2\pi)^3 \delta (\vec{k} + \vec{k}') P_\delta(k, \eta, \eta') ,
\end{equation}
we arrive at
\begin{align} \label{power_spect_kappa}
\begin{aligned}
&\expval{\kappa(\eta_i, \eta_j; \hat{n}) \kappa(\eta_l, \eta_m; \hat{n}')}= \frac{9 H_0^4 \Omega^2_{\text{m},0}}{4}\int \dd{\eta} \int \dd{\eta'} \\
&\times q_{ij}(\eta) q_{lm}(\eta') \int \frac{\dd[3]{k}}{(2\pi)^3} P_\delta(k, \eta, \eta') e^{-i \vec{k} (\eta\hat{n} - \eta'\hat{n}')}   .
\end{aligned}
\end{align}

The typical angular separation of multiply-lensed sources in galaxy lensing campaigns is in the ballpark of arcseconds. This means that the proper transverse distance between the relevant geodesics is smaller than $\sim10$~kpc, which is a small separation w.r.t LSS. In the following, we will therefore compute the cosmological correlators at the same line of sight, $\hat n=\hat n'$. 
With this simplification, the integral for the variance of differential convergence reads
\begin{align} \label{full_differential_kappa}
\begin{aligned}
&\expval{(\kappa(\eta_l, \eta_m; \hat{n}) - \kappa(\eta_n, \eta_o; \hat{n}))^2} = \frac{9 H_0^4 \Omega^2_{\text{m},0}}{2(2\pi)^2}\int \dd{\eta} \int \dd{\eta'} \\
&\times q_{lmno} (\eta, \eta') \int \dd{k} k^2 j_0(k(\eta-\eta')) P_\delta(k, \eta, \eta')  ,
\end{aligned}
\end{align}
with 
\begin{align}
\begin{aligned}
q_{lmno} (\eta, \eta') &:= q_{lm}(\eta) q_{lm}(\eta') + q_{no}(\eta) q_{no}(\eta') \\
&- 2 q_{lm}(\eta)q_{no}(\eta') .
\end{aligned} 
\end{align}
The quantities we are mostly interested  in are
\begin{align}
\begin{aligned}
&\expval{(\delta\kappa^{\rm s}_{ij})^2} =     \expval{(\kappa(\eta_i, 0; \hat{n}) - \kappa(\eta_j, 0; \hat{n}))^2} ,
\end{aligned}
\end{align}
and 
\begin{align}
\begin{aligned} \label{kappa_variance}
&\expval{(\kappa^{\rm s})^2} = \frac{9 H_0^4 \Omega^2_{\text{m},0}}{2(2\pi)^2}\int \dd{\eta} \int \dd{\eta'} \\
&\times q_{\rm s o s o} (\eta, \eta') \int \dd{k} k^2 j_0(k(\eta-\eta')) P_\delta(k, \eta, \eta').  
\end{aligned}
\end{align}
(We remind the reader that the indices o,s denote observer, source respectively.)
Analogous formulas hold for $\expval{(\delta\kappa^{\rm ls}_{ij})^2}$ and $\expval{(\kappa^{\rm ls})^2}$, $\expval{(\kappa^{\rm l})^2} $.
The line of sight integrals invoke the power spectrum of matter density perturbations $\delta$, computed at non-equal times $\eta,\eta'$. In  Sec.~\ref{sec:compute_kappa} we estimate these correlators using HALOFIT~\cite{Peacock:2014,Takahashi:2012em}. Our numerical results, obtained through this computation, are illustrated in Fig.~\ref{fig:kappa_ls_s_l}. 

\subsection{Evaluation using HALOFIT} \label{sec:compute_kappa}
The main difficulty in evaluating expressions for the variance of the external convergence is obtaining a reliable estimate of the non-equal time matter power spectrum $P_\delta(k, \eta, \eta')$\footnote{We note that simply neglecting the unequal time contribution to the correlator can bring biases when discussing projection fields like external convergence~\cite{Kitching:2016xcl}.}. This problem has been extensively studied in the literature, both analytically and numerically (see for instance~\cite{Keeton:1996tq,Holder:2002hq,Dalal:2004as,Momcheva:2005ex,Suyu:2009by} and references therein). The purpose of this section is to provide a simple, yet accurate enough analytical approximation to $P_\delta(k, \eta, \eta')$, which can be used to easily estimate the typical magnitude of external convergence given the lens and sources configuration.

In linear theory, the non-equal time matter power spectrum is simply given by 
\begin{equation}
\label{P_linear}
P_\delta(k, \eta, \eta') = D(\eta)D(\eta') P_{\rm lin}(k) \;,
\end{equation}
where $P_{\rm lin}(k)$ is the liner power spectrum evaluated at redshift zero and $D(\eta)$ is the linear theory growth factor. However, since a significant contribution to the external convergence comes from very nonlinear scales, the linear theory estimate is not reliable. Indeed, as we are going to see making comparison to the results from simulations with ray-tracing, the linear theory predictions significantly underestimate the variance of external convergence. 

To get a more reliable theoretical estimate, one has to use the nonlinear matter power spectrum, which can be simply obtained using HALOFIT~\cite{Peacock:2014,Takahashi:2012em}. Unfortunately, HALOFIT outputs the nonlinear power spectrum only at equal times. To extend this output to non-equal times requires some approximations. Inspired by the linear theory, the commonly used prescription is
\begin{equation}
\label{eq:mid-point-approximation}
P_\delta(k, \eta, \eta') = \sqrt{P_\delta(k, \eta)P_\delta(k, \eta')} \;.
\end{equation}
We are going to argue that this and other similar approximations do not properly capture the non-equal time matter power spectrum on small scales. The reason is large bulk flows, which displace the dark matter particles by $\mathcal O(10)$~Mpc. These large displacements exactly cancel for equal time correlation functions\footnote{Large displacements can have observable effects only for sharp features in the correlation functions. Baryon acoustic oscillation (BAO) peak is one such feature and large displacements lead to the spread of the BAO peak, or damping of the BAO wiggles in the power spectrum.}, due to the equivalence principle. However, for non-equal time correlation functions, the dark matter particles are displaced by different amounts, depending on times at which the density fields are evaluated. On scales smaller than $\mathcal O(10)$~Mpc, this leads to exponential suppression of power in the non-equal time power spectrum. This important effect is not captured by Eq.~\eqref{eq:mid-point-approximation}.

In order to gain some intuition about how large displacements affect the non-equal time power spectrum, we can use Lagrangian perturbation theory. In this setup we have (with $ \vec{r} $ the Euclidean coordinate and $ \vec{q} $ the Lagrangian coordinate)
\begin{equation}
1 + \delta(\vec{r}) = \int \dd[3]{q} \delta(\vec{r} - \vec{q} - \vec{\psi}(\vec{q}))  ,
\end{equation}
where $ \vec{\psi} $ is the displacement field
\begin{equation}
\vec{r} (\vec{q}, \eta) = \vec{q} + \vec{\psi}(\vec{q} , \eta )  .
\end{equation}
In Fourier space,
\begin{equation}
\delta(\vec{k}) = \int \dd[3]{q} \e^{-\iu \vec{k} (\vec{q} + \vec{\psi})} \ .
\end{equation}
Hence, we can write the two-point correlator as
\begin{align}
\begin{aligned}
\expval{\delta(\vec{k}, z_1) \delta(\vec{k}', z_2)} &= \int \dd[3]{q_1}\int \dd[3]{q_2} \\ 
\times&\expval{\e^{-\iu \vec{k} (\vec{q}_1 + \vec{\psi}_1)}  \e^{-\iu \vec{k}' (\vec{q}_2 + \vec{\psi}_2)}  }  ,
\end{aligned}
\end{align}
where we used the shorthand $ \vec{\psi}_i := \vec{\psi}(\vec{q}_i, z_i) $. Using homogeneity and isotropy of the universe, we can write 
\begin{align}
\begin{aligned}
& \expval{\delta(\vec{k}, z_1) \delta(\vec{k}', z_2)} = (2\pi)^3 \delta(\vec{k} + \vec{k}') \\ 
& \qquad \qquad \times \int \dd[3]{q}  \e^{-\iu \vec{q} \cdot \vec{k}}  \expval{\e^{-\iu \vec{k} (\vec{\psi}(\vec{q},z_2)  - \vec{\psi}(0,z_1)) }}  ; 
\end{aligned}
\end{align}
which translates into the following formula for the non-equal time power spectrum
\begin{equation}
P_\delta(k, \eta, \eta') = \int \dd[3]{q} \e^{-\iu \vec{q} \cdot \vec{k}} \expval{\e^{-\iu \vec{k} (\vec{\psi}(\vec{q},\eta')  - \vec{\psi}(0,\eta)) }}  .
\end{equation}
Note that for two different times the relative displacement in the exponent can be large. For a large $k$ (small scales) this implies that the contribution to the power spectrum becomes exponentially suppressed, as we argued at the beginning of this section. 

We can calculate this exponential suppression a bit more explicitly. For this purpose we can focus on the simplest case of Zel'dovich approximation. The Zel'dovich displacement is simply given in terms of the linear density field as follows
\begin{equation}
\vec{\psi}_{\rm Z} (\vec{q},\eta) = \int \frac{\dd[3]{k}}{(2\pi)^3} \e^{i\vec{k}\cdot \vec{q}} \frac{i\vec{k}}{k^2} \delta_{\rm lin}(\vec{k},\eta) \;.
\end{equation}
Using the cumulant theorem and assuming Gaussian initial conditions, the non-equal time Zel'dovich power spectrum is given by
\begin{equation}
P_{\rm Z}(k,\eta,\eta') = \e^{-k^2\Sigma^2 (D(\eta)-D(\eta'))^2/2} P_{\rm Z}(k,\bar\eta) \;,
\end{equation}
where 
\begin{equation}
\Sigma^2=\frac{1}{6\pi^2} \int_0^\infty \dd{k} P_{\rm lin} (k,0) \;,
\end{equation}
$P_{\rm Z}(k,\eta)$ is the standard equal-time Zel'dovich power spectrum and we have defined $D(\eta_1) D(\eta_2) =: D^2(\bar{\eta}) $, with $\bar{\eta}$ an appropriate mean comoving distance which can be determined using the form of the linear growth factor $ D $. The same result was obtained in~\cite{Chisari:2019tig} (for a similar discussion see also~\cite{Zhang:2021uyp}). 

One can show that the same exponential suppression remains going to higher orders in perturbation theory. However, beyond Zel'dovich approximation, the nonlinear spectra cannot be simply expressed through the equal time counterparts anymore. For instance, the general structure of the one-loop result can be written as
\begin{align}
\begin{aligned}
P_{\rm 1-loop}(k,\eta,\eta') &= \e^{-k^2\Sigma^2 (D(\eta)-D(\eta'))^2/2} \\
& \times [P_{\rm 1-loop}(k,\bar\eta) + \delta P (k,\eta,\eta') ] \;.
\end{aligned}
\end{align}
The exact form of $\delta P (k,\eta,\eta')$ is not important, but we know that it has two important properties. First, this correction is small in perturbation theory~\cite{Vlah:2016bcl,Schmittfull:2018yuk}. Second, $\delta P (k,\eta,\eta')$ vanishes for equal times. Therefore, we expect that the correction to the equal-time one-loop term in the square brackets is always small. Furthermore, given the expectation that $P (k,\eta,\eta')$ is a smooth function of $\eta$ and $\eta'$, when the two times are not equal, the exponential suppression at high $k$ is always large enough to make any small mistake in the modeling of the nonlinear power spectrum insignificant. 

Motivated by these results, we make the following ansatz for the non-equal time power spectrum
\begin{equation}
\label{P_Castorina}
P_\delta(k,\eta,\eta') = \e^{-k^2\Sigma^2 (D(\eta)-D(\eta'))^2/2} P_\delta(k,\bar\eta) \;.
\end{equation}
This equation has the correct equal-time limit, on large scales (small $k$) it reduces to the linear theory result given by Eq.~\eqref{P_linear}, and on small scales it has the correct exponential suppression of power induced by the difference in magnitudes of large bulk flows at different redshift. The equal time power spectrum on the right hand side $P_\delta(k, \bar{\eta})$ can be simply evaluated using the  HALOFIT~\cite{Peacock:2014,Takahashi:2012em}. Eq.~\eqref{P_Castorina} can be used in Eq.~\eqref{full_differential_kappa} to compute the differential external convergence variance. The highly oscillating integrand (due to the presence of the Bessel function $j_0$) can be tamed by means of FFTlog techniques~\cite{Hamilton:1999uv,Schoneberg:2018fis}. Finally, we introduce a cut-off in the $k$ integral at some $k_{\rm cutoff}$. We choose $k_{\rm cutoff}=10~{\rm Mpc}^{-1}$, where individual galaxies and baryonic effects most likely lead to a breakdown of the HALOFIT result. Note that similar smoothing is implicitly used in the ray-tracing simulations when the gravitational potential is estimated from the distribution of matter. Changing $k_{\rm cutoff}$ by a factor of 2 up or down affects our results at the level of a few tens of percent, which is comparable to other theoretical uncertainties in our equations. In Sec.~\ref{a:compNbodPT} we compare the results of our calculations with results obtained in the literature using ray tracing techniques.

\subsection{Comparison to ray-tracing results}\label{a:compNbodPT}
TDCOSMO derives Bayesian priors for external convergence by using ray-tracing through the Millennium simulation~\cite{Springel:2005}, on LOSs which are chosen to match the galaxy density observed in each strong lensing system of interest~\cite{Hilbert:2008kb,Suyu:2009by,Suyu:2012aa}. We can use these numerical results to compare with our analysis (Eq.~\eqref{kappa_variance}).

\begin{figure*}
    \centering
    \includegraphics[scale=0.36]{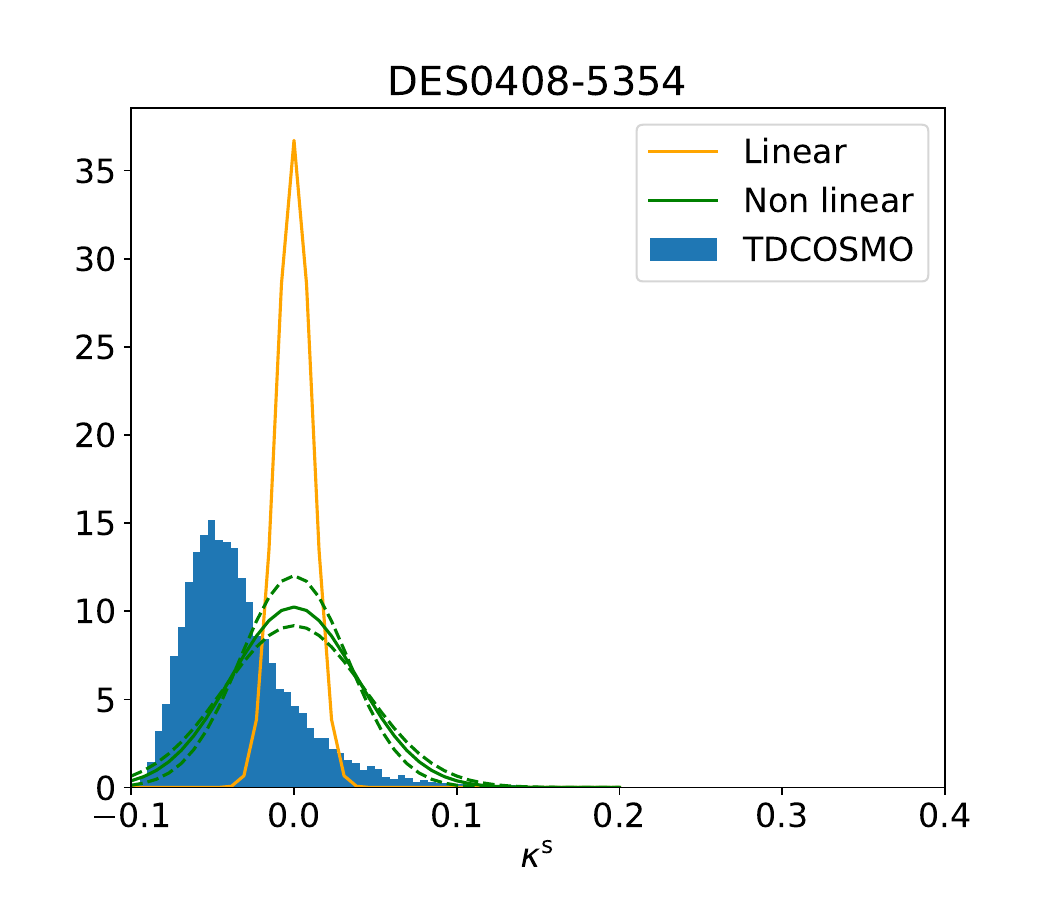}
    \includegraphics[scale=0.36]{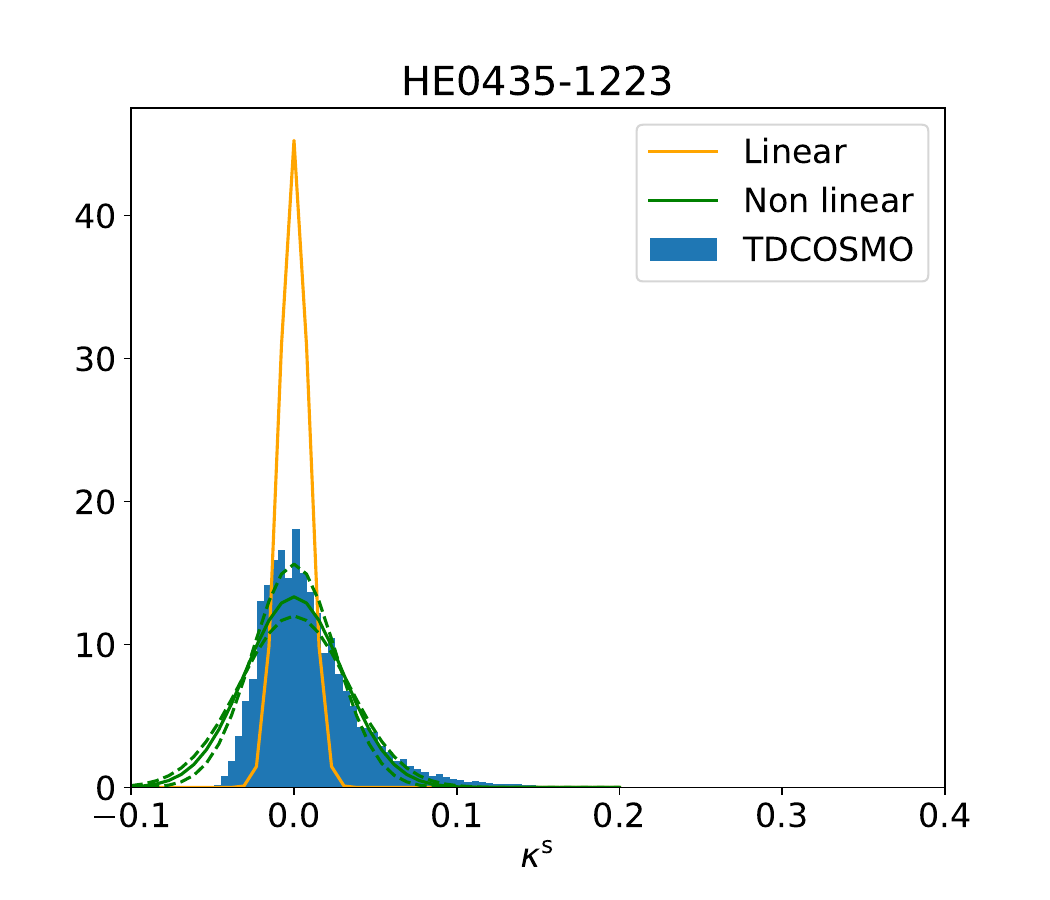}
    \includegraphics[scale=0.36]{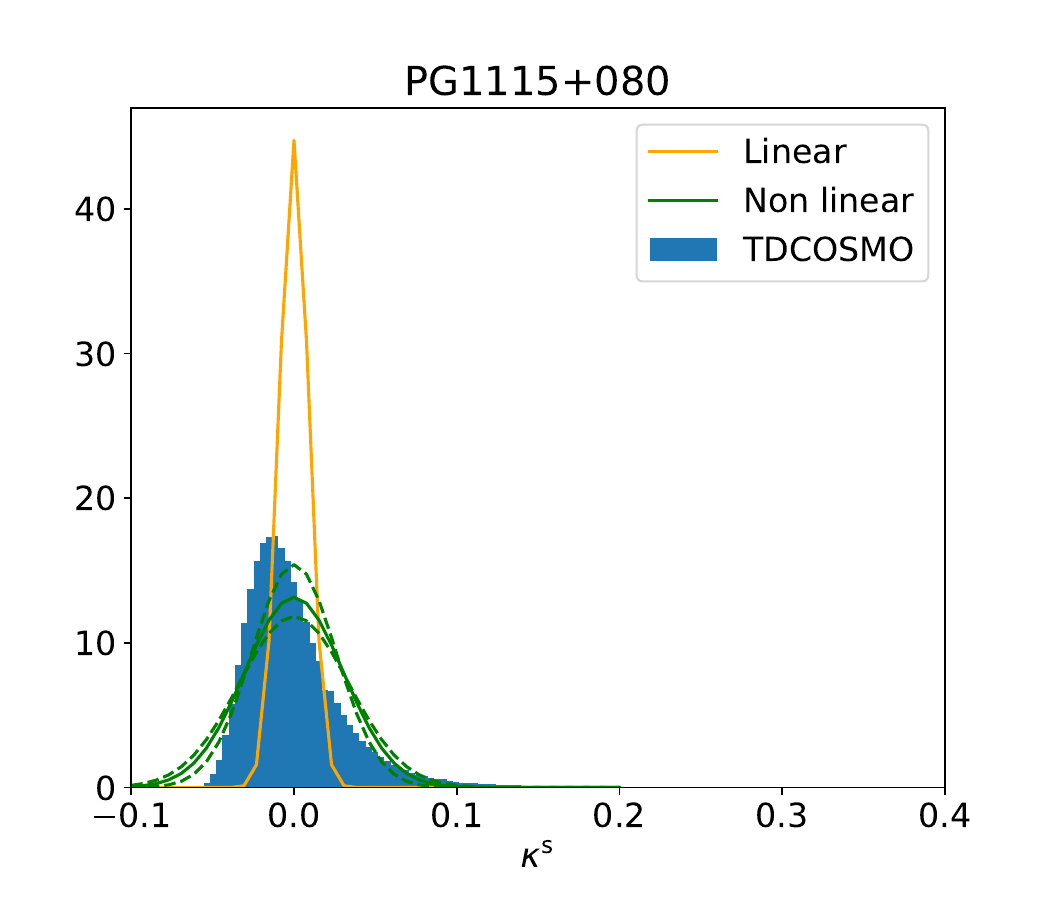}
    \includegraphics[scale=0.36]{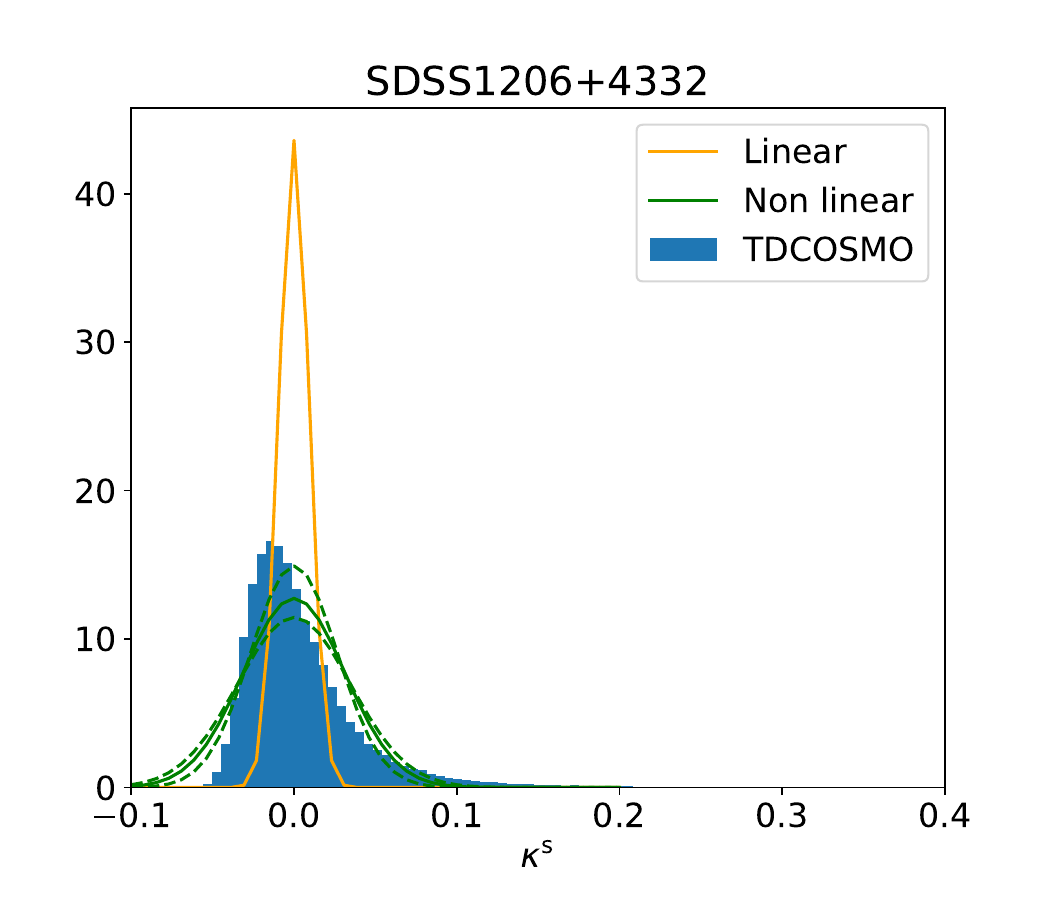}
        \includegraphics[scale=0.36]{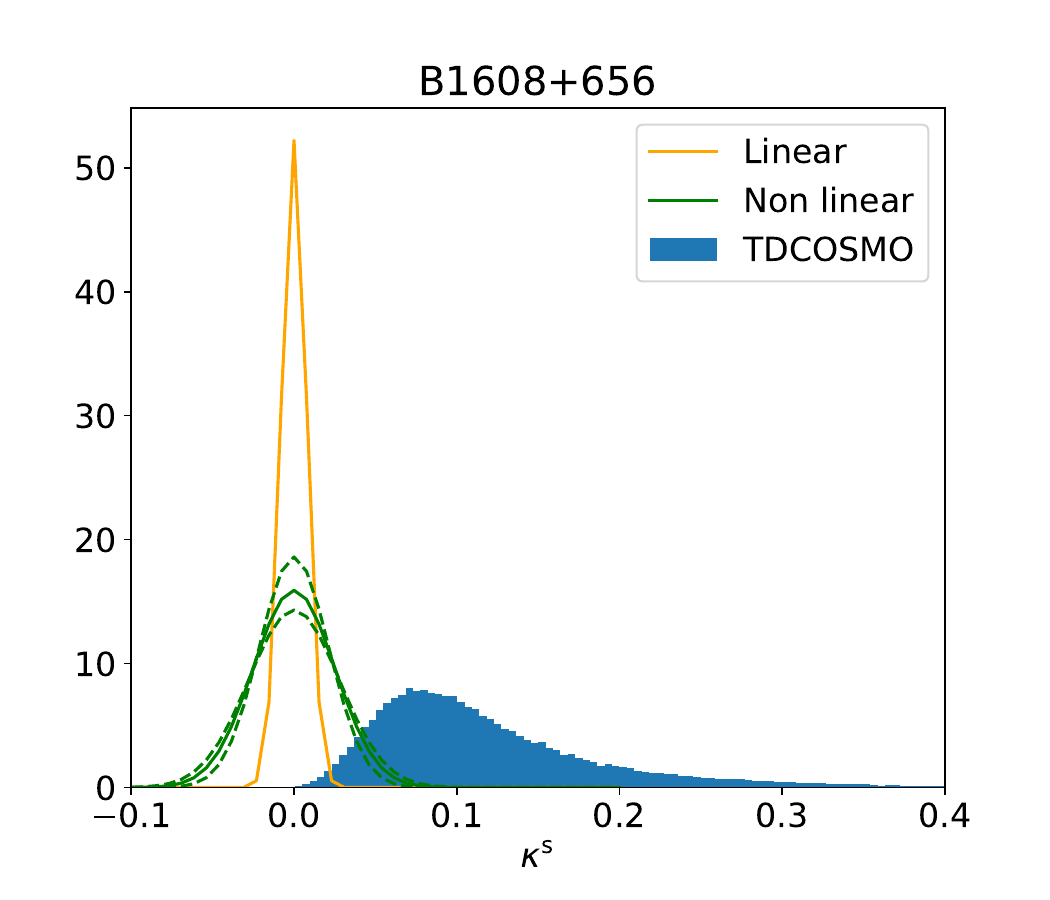}
            \includegraphics[scale=0.36]{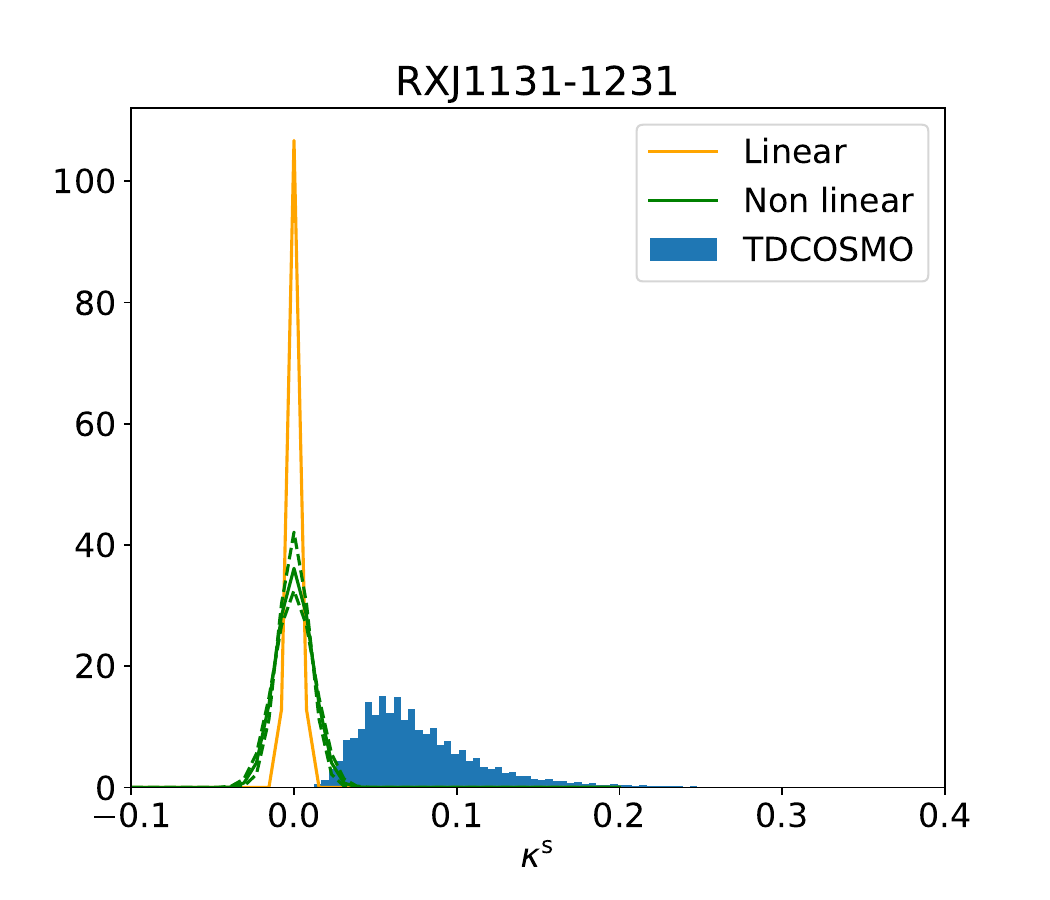}
                        \includegraphics[scale=0.36]{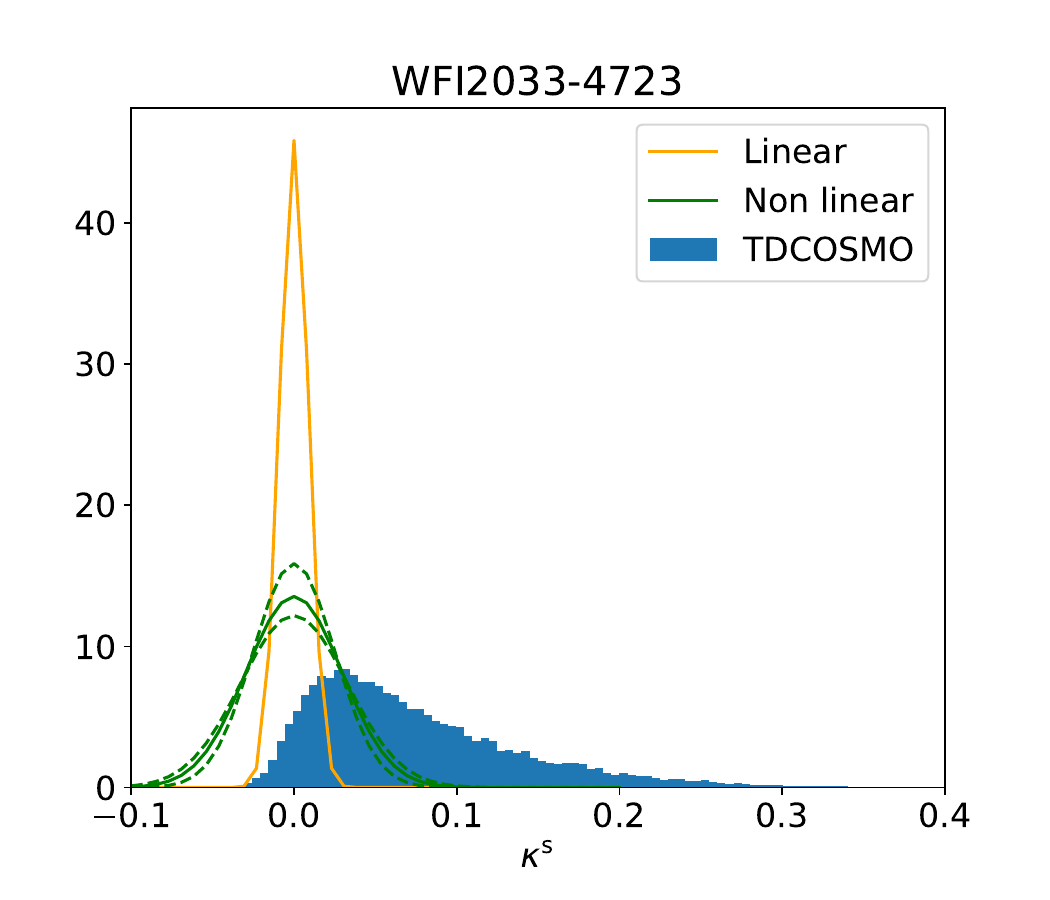}
    \caption{Comparing the probability distribution obtained in ray tracing~\cite{Suyu:2009by} (blue bar histograms) with our computation, in linear theory (solid orange) and with the non-linear approximation (solid green: $k_{\rm cutoff}=10~{\rm Mpc}^{-1}$, dashed green: $k_{\rm cutoff}=5\,{\rm and}\,20~{\rm Mpc}^{-1}$). We remark that our results cannot reproduce the bias on the external convergence (nonzero mean seen in some of the blue bar histograms), since our computation is equivalent to an average over all LOSs, differently from the ray-tracing analysis TDCOSMO performs, which is calibrated to match the richness of the actual lensing systems. A fairer comparison between our computation and typical TDCOSMO results, obtained by averaging over many LOSs, is shown in Fig.~\ref{fig:all_Los}. Code: \href{https://github.com/lucateo/Comments_MSD/blob/main/Notebooks/delta_kappa_nonlinear.ipynb}{\faGithub}.}
    \label{fig:comparison}
\end{figure*}

\begin{figure}
    \centering
    \includegraphics[scale=0.5]{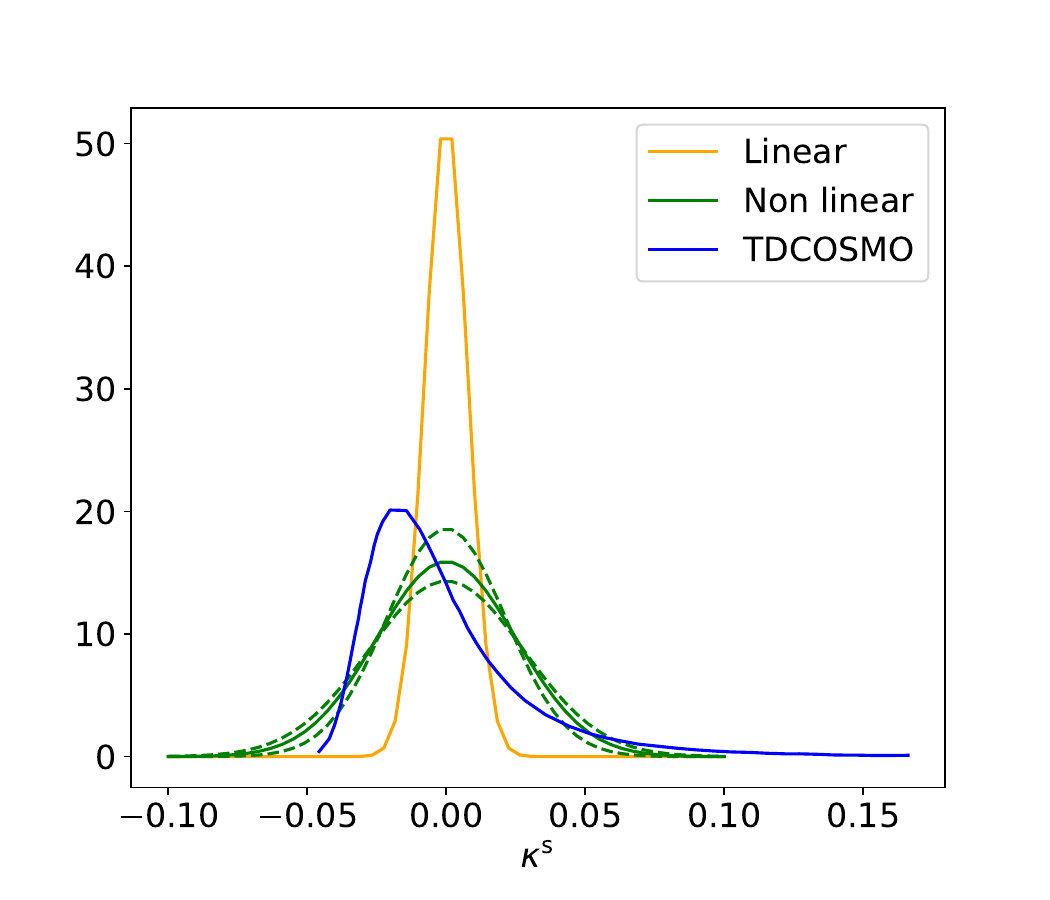}
    \caption{Comparing the distribution of the system B1608+656, when the average in the Millennium Simulation is done over all LOSs, with our estimates. Code: \href{https://github.com/lucateo/Comments_MSD/blob/main/Notebooks/delta_kappa_nonlinear.ipynb}{\faGithub}.}
    \label{fig:all_Los}
\end{figure}

In Fig.~\ref{fig:comparison} and Tab.~\ref{tab:kappa_table} we compare our computation (linear, obtained using Eq.~\eqref{P_linear} for the power spectrum; and non-linear, obtained using Eq.~\eqref{P_Castorina}) with ray tracing results from the TDCOSMO collaboration available \href{https://github.com/TDCOSMO/hierarchy_analysis_2020_public/tree/master/TDCOSMO_sample/TDCOSMO_data}{\emph{here}}. Fig.~\ref{fig:all_Los} shows the results obtained in Ref.~\cite{Suyu:2009by} for the probability distribution of external convergence, averaging over all LOSs (that is, not restricting to fields containing strong lensing systems). 

Our nonlinear analysis (incorporating HALOFIT and the non-equal time approximation) reproduces the variance in $\kappa^{\rm s}$ to within about 30\% accuracy for the systems which have a mean value of $\kappa^{\rm s}$ compatible with zero (first 4 systems in Tab.~\ref{tab:kappa_table}, top 4 panels in Fig.~\ref{fig:comparison}). Some systems, however, are found in~\cite{Suyu:2009by} to be biased with a mean $\kappa^{\rm s}$ that is significantly off zero (last 3 systems in Tab.~\ref{tab:kappa_table}, bottom 3 panels in Fig.~\ref{fig:comparison}). This probably reflects excess structure along the LOS, typical of systems in crowded fields. For these systems, our calculation not only misses the bias, but also underestimates the spread in $\kappa^{\rm s}$, by up to a factor of $\sim4$. Thus, indeed, the simplified computation from the previous section can only be used to provide a rough estimate of the magnitude of weak lensing effects, and ray tracing analyses on the lines of Refs.~\cite{Suyu:2009by, Tihhonova:2017mym,Fassnacht:2005uc,Greene:2013xoa} are probably mandatory on a system by system study.
\begin{table}
	\centering
	\begin{tabular}{ccccc}
		\toprule
System & $\sigma_{\rm lin}$  & $\sigma_{\rm halofit}$  & $\sigma^{\rm TDCOSMO}$ & $ \kappa_{\rm ext}^{\rm TDCOSMO} $  \\
\midrule
DES0408-5354 & $0.0109$ & $0.0390$ & $0.0380$ & $-0.0397^{+0.0421}_{-0.0242}$ \\
HE0435-1223 & $0.0088$ & $0.0299$ & $0.0342$ & $0.0040^{+0.0363}_{-0.0215}$ \\
PG1115+080 & $0.0089$ & $0.0303$ & $0.0330$ & $-0.0054^{+0.0358}_{-0.0209}$ \\
SDSS1206+4332 & $0.0092$ & $0.0313$ & $0.0410$ & $-0.0037^{+0.0402}_{-0.0215}$ \\
B1608+656 & $0.0076$ & $0.0251$ & $0.0903$ & $0.1026^{+0.0949}_{-0.0451}$ \\
RXJ1131-1231 & $0.0037$ & $0.0110$ & $0.0433$ & $0.0695^{+0.0480}_{-0.0260}$ \\
WFI2033-4723 & $0.0087$ & $0.0295$ & $0.0660$ & $0.0591^{+0.0863}_{-0.0442}$ \\
		\bottomrule
	\end{tabular}
	\caption{Comparing our external convergence estimates with ray tracing results from the literature. We show the external convergence variance using linear theory (Eq.~\eqref{P_linear}) on the second column and non-linear approximation (Eq.~\eqref{P_Castorina}) on the third column. The ray tracing results from the TDCOSMO collaboration (available \href{https://github.com/TDCOSMO/hierarchy_analysis_2020_public/tree/master/TDCOSMO_sample/TDCOSMO_data}{\emph{here}}) are shown in the last two columns ($\kappa^{\rm s}$ variance on the fourth, $\kappa^{\rm s}$ mean and the 16th and 86th percent quantiles on the fifth).}
	\label{tab:kappa_table}
\end{table}

\vspace{6 pt}

\bibliography{ref}

\end{document}